\documentstyle[11pt,citesort]{article}
\input epsf
\def \is {\! & \! = \! & \! }
\newcommand{\newsubsection}[1]{
\vspace{1cm}
\pagebreak[3]
\addtocounter{subsection}{1}
\noindent{\large\bf 
\thesubsection.
#1}
\nopagebreak
\vspace{2mm}
\nopagebreak}

\renewcommand{\thesubsection}{\arabic{subsection}}

\newcommand{\figuur}[3]{
\begin{figure}[t]\begin{center}
\leavevmode\hbox{\epsfxsize=#2 \epsffile{#1.eps}}\\[3mm]
\parbox{15.5cm}{\small
\it #3}
\end{center}
\end{figure}}
\newlength{\extraspace}
\newlength{\extraspaces}
\newcommand{\TT}{{\mbox{\small $T$}}}
\newcommand{\XX}{{\mbox{\small $X$}}}
\newcommand{\ba}{\begin{eqnarray}
\addtolength{\abovedisplayskip}{\extraspaces}
\addtolength{\belowdisplayskip}{\extraspaces}
\addtolength{\abovedisplayshortskip}{\extraspace}
\addtolength{\belowdisplayshortskip}{\extraspace}}
\newcommand{\ea}{\end{eqnarray}}
\newcommand{\be}{\begin{equation}
\addtolength{\abovedisplayskip}{\extraspaces}
\addtolength{\belowdisplayskip}{\extraspaces}
\addtolength{\abovedisplayshortskip}{\extraspace}
\addtolength{\belowdisplayshortskip}{\extraspace}}
\newcommand{\ee}{\end{equation}}

\begin{document}
\addtolength{\baselineskip}{.8mm}

\def \two {{{}_{2}}}
\def \TTT{{\bf T}}

\newcommand{\ggt}{\!>\!\!>\!}
\newcommand{\llt}{\!<\!\!<\!}
\def\dag{\dagger}
\def\bea{\begin{eqnarray}}
\def\eea{\end{eqnarray}}
\newcommand{\eff}{{\rm eff}}
\newcommand{\tr}{{\rm tr~}}
\newcommand{\Tr}{{\rm tr~}}
\def\bra#1{{\langle}#1|}
\def\ket#1{|#1\rangle}
\def\vev#1{\langle{#1}\rangle}
\def\CA{{\cal A}}
\def\CC{{\cal C}}
\def\sect#1{\bigskip\noindent{\bf #1}\bigskip}
\def\IR{\relax{\rm I\kern-.18em R}}

\begin{titlepage}
\begin{center}

{\hbox to\hsize{ \hfill PUPT-2083}}
{\hbox to\hsize{ \hfill hep-th/0304224}}

\bigskip

\vspace{3\baselineskip}

{\LARGE \bf Strings from Tachyons:}\\[6mm]
{\Large \bf The {\bf \sc $c=1$} Matrix Reloaded}

\bigskip
\bigskip
\bigskip

{\large John McGreevy and Herman Verlinde }\\[1.5cm]

{\large \it Department of Physics, Princeton University,
Princeton,
NJ 08544}\\[6mm]

\vspace*{1.5cm}

{\bf Abstract}\\

\end{center}
We propose a new interpretation of the $c\!=\!1$ matrix model as
the world-line theory of $N$ unstable D-particles, in which the
hermitian matrix is provided by the non-abelian open string
tachyon. For D-branes in 1+1-d string theory, we find a direct
quantitative match between the closed string emission due to a
rolling tachyon and that due to a rolling eigenvalue in the matrix
model. We explain the origin of the double-scaling limit, and
interpret it as an extreme representative of a large equivalence
class of dual theories. Finally, we define a concrete decoupling
limit of unstable D-particles in IIB string theory that reduces to
the $c\! =\! 1$ matrix model, suggesting that 1+1-d string theory
represents the near-horizon limit of an ultra-dense gas of IIB
D-particles.

\end{titlepage}

\newsubsection{Introduction}

The duality between open and closed string theory has led to
various deep and surprising insights into the fundamental
structure of both systems. A relatively recent, but still
incompletely understood, example of such a duality is the
observation that unstable D-branes can completely decay into
closed strings via open string tachyon condensation \cite{piljin1}
\cite{Sen:99}. Both the full time evolution as well as the final
stage of this decay process are fascinating arenas for further
study
\cite{Sen:2002nu}\cite{Sen:2002vv}\cite{piljin2}\cite{Gutperle:2002ai}\cite{Gutperle:2003xf}.
In particular, it would be desirable to find a controllable
description of closed string creation from open string tachyon
matter \cite{stherm}\cite{Lambert:2003zr}\cite{GIR}.

It is a natural strategy to try to apply the lessons of other open/closed
string dualities, like Matrix theory and  the AdS/CFT
correspondence, to this problem. In particular, we could attempt to
find a regime in which the tachyon degrees of freedom are
naively decoupled from the bulk closed strings, but at the same
time become fully equivalent to a complete closed string theory in
an appropriate near horizon geometry. Experience tells us that
this can be achieved if we can take a suitable large $N$ limit and
tune parameters, such that the tachyon matter becomes ultra-light
and saturates all possible degrees of freedom of the theory.

With this motivation, we will consider in this paper the
non-abelian tachyon dynamics of a dense gas of many unstable
D-particles. To enable investigation of the tachyon mode in
isolation, we will consider special models, or regimes of
couplings, in which all other degrees of freedom of the
D-particles, such as their space-time positions, are either absent
or decoupled. A specific string model with this property is
1+1-dimensional bosonic string theory, which has a well-known dual
description in terms of matrix quantum mechanics;
for reviews see {\it e.g.}
\cite{Ginsparg:is}\cite{Klebanov:1991qa}\cite{Polchinski:1994mb}\cite{matrix}.

This duality between 1+1-d strings and $c=1$ matrix quantum
mechanics is the oldest known example of a holographic
equivalence, and has several attractive features in comparison
with the examples found later.
It is a holographic theory with an
$S$-matrix description, and therefore more similar to a model of
holography in flat space. Secondly, both sides of the duality have
overlapping weakly coupled regimes; the matrix model is even
exactly soluble. This allows very precise quantitative
comparisons. The duality also has various mysterious features and
unresolved puzzles \cite{Polchinski:1994jp}.

In the light of more recent developments, it is natural to suspect
that the $c=1$ matrix degree of freedom should be related to the
open string tachyon of unstable D-particles of the 1+1-d string
theory itself.\footnote{Indeed, a comment to this effect appears
in \cite{Polyakov:2001af}.} We will present concrete evidence in
support of this identification. In particular we will find a
direct quantitative match between the closed string emission due
to a rolling tachyon and that due to a rolling eigenvalue in the
matrix model. Via this new interpretation of the matrix model, we
will be able to clarify several of its somewhat mysterious
features. We will explain the physical meaning of the
double-scaling limit as a decoupling limit, and interpret it as
selecting an extreme representative of a large equivalence class
of dual theories in which D-branes are replaced by their
back-reaction on the closed string background. The projection onto
singlet states is naturally implemented by the worldline gauge
invariance. The D-brane perspective also sheds new light on the
non-perturbative instability of the matrix model against tunneling
of eigenvalues towards the wrong side of the potential barrier: it
corresponds to decay of the open string tachyon towards the regime
where its potential is unbounded from below. Due to this
instability, it would appear that 1+1-d string theory is an
incomplete model, since it does not seem to have a completely
self-consistent non-perturbative definition.

It is an important question, therefore, whether it is possible to
obtain 1+1-d string theory via a special limit of one of the
consistent supersymmetric string theories. In the last section, we
will propose a confirmative answer to this question, by defining a
natural decoupling limit of a dense collection of unstable
D-particles in IIB string theory, in which the world-line theory
reduces to the $c=1$ matrix model. This correspondence suggests
that 1+1-d string theory can be given its rightful place within
the world of consistent theories, as the near-horizon limit of a
dense cluster of unstable D-particles.

\renewcommand{\footnotesize}{\small}

\newsubsection{Rolling and Bouncing Tachyons}

We begin with a brief summary of some recent results and insights
about open string tachyon dynamics on non-supersymmetric D-branes,
that will be useful for our later discussion. We will restrict our
attention to the case of unstable D-particles, both in bosonic and
supersymmetric string theory.

The worldline theory of a D-particle in bosonic string theory is a
quantum mechanical system, with one unstable tachyonic degree of
freedom $T$. Its equation of motion is the requirement that the
corresponding worldsheet boundary interaction \be S_{open} = \int
\!\! d\xi \; T(\XX^0(\xi))\, \ee defines a proper boundary
conformal field theory. In case the closed string background is
static, so that \be S_{closed} = S_{CFT} + {1\over 4\pi} \int \!\!
d^2\sigma \, (\partial_\alpha \XX^0)^2, \ee where $S_{CFT}$ is any
$c=25$ CFT describing the spatial directions of the target space,
the following trajectories \be \label{roll}
T_{\rm roll}(X^0) = \lambda \exp
X^0 \ee \be \label{bounce} T_{\rm bounce}(\XX^0) = \lambda \cosh
\XX^0 \ee are exact solutions for all values of $\lambda \leq
1/2$. The first trajectory describes a rolling solution: the
tachyon starts at the top of the potential and rolls towards the
minimum  of its potential at $T=\infty$. The second trajectory
represents a bounce: the tachyon starts from and returns to the
minimum at $T=\infty$, reaching its smallest value $T=\lambda$ at
$\XX^0=0$. In the interacting string theory, this bouncing tachyon
solution can be thought of as being initiated at early times by a
collision of closed string matter, creating an unstable D-particle
which, after a finite lifetime of order $|\log \lambda|$ (for
small $\lambda$), decays back into closed strings. The critical
configuration with $\lambda=1/2$ has the special feature that the
boundary state formally vanishes \cite{Sen:2002nu}; it can be shown to be
equivalent to a specific time-dependent classical closed string
background \cite{stherm}. This equivalence can be thought of as a perturbative
(or infinitesimal) version of open/closed string duality, since it
indicates that a D-particle with a particular tachyon profile can
be completely absorbed via an adjustment of the closed string
background. This surprising result will be a key element in our
later argumentation.

The classical behavior (\ref{bounce}) of the
tachyon mode can be reproduced in terms of
an effective mechanical model, with a Born-Infeld type lagrangian
\footnote{
Following 
\cite{Lambert:2003zr},\cite{kutasov:2003er} 
we have performed a field redefinition relative to (\ref{bounce});
the variable $\TT$ appearing below has the solution
$ \sinh^2 T/2 = \lambda~\cosh t$.}
\be
\label{eff}
S(\TT) = - \int\!\!dt\, V(\TT)
\sqrt{1-\dot{\TT}^2}
\ee
where
\be
\label{special}
V(\TT)
= {1
\over g_s \cosh({\TT/2})}.
\ee
This form of $V(\TT)$ applies for positive $T$ only:
for negative $T$, the bosonic string tachyon potential is known to
be unbounded from below. We will for the most part restrict our
attention to the stable region $T>0$.
In the following section we will use the matrix generalization of this
effective model as our starting point for studying the quantum mechanics
of the matrix valued tachyon associated with
systems of many unstable
D-particles. The detailed analytic form of the potential $V(T)$ will not
be essential, except for the two global properties that (i) it has a
single maximum at $T=0$, near which it behaves as
\be
\label{approx}
V(T)\simeq {1\over g_s}\,  \Bigl(1 - {1\over 8}\, T^2\,  + \ldots \Bigr),
\ee
and (ii) $V(T)$ exponentially decays to zero at $\TT\rightarrow \infty$.

In a general closed string background, the unstable D-particle has
many more light degrees of freedom than just $T$; in particular it
has coordinates $X^i(t)$ that parametrize the space-time motion of
the D-particle. To enable investigation of the tachyon mode in
isolation, we will consider special models, or regimes of
couplings, in which these other degrees of freedom are either
absent or decoupled. A specific bosonic string model which is
known to satisfy this property is the 1-dimensional non-critical
string, or equivalently, the 1+1-dimensional critical string
theory, to which we now turn.

\newsubsection{D-branes in 1+1-d String Theory}

The space-like motion of the 1+1-d critical string is
described by the Liouville conformal field theory
\be S_{{\rm bulk}} = {1 \over 4 \pi} \int
d^2 \sigma \left((\partial_a \varphi)^2 + Q R^{(2)} \varphi + 4
\pi \mu e^{2 b \varphi} \right), \ee
with $Q = b + {1 \over b}$ and central charge $ c_L = 1 + 6 Q^2 .$
The case of interest to us is $ c_L = 25,~ Q = 2,~ b=1 ,$ but it
will sometimes be useful to keep $b \neq 1$ as a
regulator.
The $c_L =25$ Liouville CFT represents a classical string background
of the effective 1+1-d target space-time field theory (here
${\cal T}$ denotes the {\it closed} string tachyon, and $R$
the 1+1-d target-space curvature scalar)
\be
\label{target}
S_{\eff} = \int\!\!
d^2 x\, \sqrt{-G}e^{-2\Phi}\Bigl(R + 4(\nabla \Phi)^2 - (\nabla
{\cal T}\, )^2 + 4 {\cal T}\, {}^2 + 16 + \ldots \Bigr). \ee
Besides the standard classical tachyon profile ${\cal T} \simeq e^{2\varphi}$,
this action also admits  ${\cal T} \simeq \varphi e^{2\varphi}$ as a static
solution for the closed string tachyon, and the latter solution dominates
for large negative $\varphi$. The $c\! =\! 1$ string background
\be
\Phi(\varphi) \simeq 2
\varphi \, , \qquad \qquad {\cal T}(\varphi) \simeq
(\varphi + {1\over2} \log \mu)
\mu e^{2\varphi} \ee
is characterized by just one single
parameter, which we view as related to the string coupling
$g_{s} = e^{\Phi}$ at the location of the ``tachyon wall'' (i.e. the place
where ${\cal T}$ becomes of order 1 in string units) via \be
\label{geff}
g_\eff \simeq 1/\mu, \ee
or as related to the value of ${\cal T}$ at the location
$\varphi =0$ of the ``dilaton wall'' (i.e. where the string coupling is of order 1) via
\be
\label{dwall}
{\cal T}(0)  \simeq {\mu \over 2}\log \mu.
\ee
The action (\ref{target}) has only one single propagating degree of freedom, which we can
take to be the closed string tachyon ${\cal T}$. In spite of its
name, it in fact satisfies -- due to the presence of the linear
dilaton -- a massless wave-equation. Vertex operators corresponding
to normalized states look like
\be
 V_{P} = e^{(Q+iP)\varphi }
\ee
with $P$ real and
positive\footnote{These in fact do not correspond to good local
operators, since the solution to the classical Liouville equation
in their presence implies a hyperbolic metric ({\it i.e.} a
throat) in their neighborhood
\cite{Seiberg:1990eb}\cite{Polchinski:1990mh}.}.
$V_P$ has conformal dimension
$ \Delta_P = {1\over 2} (Q^2 + P^2)$.  We will call the state with this momentum
$ \ket{v_P} $, and take it to be normalized so that $\vev{v_{P^\prime} | v_P} = \pi
\delta( P - P^\prime) .$

The possible consistent boundary conditions of Liouville CFT have
recently been studied in \cite{Fateev:2000ik},
\cite{Teschner:2000md} (see also
\cite{Rajaraman:1999hn}\cite{Ponsot:2001ng}\cite{Kostov:2002uq}).
This open Liouville theory is defined by introducing the
Weyl-invariant boundary interaction \be S_{{\rm bdy}} = {1 \over 4
\pi} \int_{\partial \Gamma} \left( {Q K \over 2 \pi} \varphi +
\mu_B e^{b \varphi} \right) d \xi, \ee where $K$ is the extrinsic
curvature, $\xi$ is a coordinate on the boundary, and $\mu_B$ is
the boundary cosmological constant.  It represents a continuous
marginal coupling of the boundary CFT. \footnote{We would like to
thank J. Teschner 
for correcting an error in an earlier version of this paper.}


An interesting quantity is the overlap of the momentum eigenstate $\ket{v_P}$
with the boundary state corresponding to the boundary action described
above. It corresponds to the one-point function of the vertex operator $V_P$
on the disk. It is given by \cite{Fateev:2000ik} \be \label{fzz}
 \vev{v_P|B_s } =
{ \hat c \over i P} ( \pi \mu \gamma(b^2 ))^{ - i P/2b} \Gamma(
1 + i b P) \Gamma( 1 + iP/b) \cos( \pi s P) , \ee
with $\hat c$ an overall normalization constant, $\gamma(x)\equiv
\Gamma(x)/\Gamma(1-x) $ and $s$ is a parameter\footnote{Note the
slight change of conventions:
$$ s = s_{FZZ} = 2 s_{T},~~~~P = 2P_{FZZ} = 2P_{T} $$
where the subscript
$FZZ$ labels quantities appearing in \cite{Fateev:2000ik}
and the subscript $T$ labels those in \cite{Teschner:2000md}.}
related to
$\mu_B$ by the relation
\be \label{star} \cosh^2 \pi b s = { \mu_B
^2 \over \mu } \sin \pi b^2 .
\ee
The boundary state with label $s$ is identical to
that with label $-s$.
This explicit expression for of $\vev{v_P|B_s}$ was
found in \cite{Fateev:2000ik} by deriving a functional equation
that it must satisfy; the parameter $s$ appearing in (\ref{star})
arises in solving this equation.
Note that, unlike for D-branes in flat space, the overlap (\ref{fzz})
has a quite non-trivial $P$ dependence and is not just a phase factor.
Specializing to the critical value of $b= 1$ gives
\be
\label{fuzzy}
 \vev{v_P|B_s } = \,
c\, e^{-i\delta(P)}\, {\pi \cos (\pi s P) \over  \sinh (\pi P)}
\ee where \be \label{leg} e^{-i\delta(P)} =( \pi \mu )^{- i P/2}
{\Gamma(1 + i P)\over \Gamma(1-iP)}.  \ee
In the $b \to 1$ limit, the right hand side of (\ref{star})
remains finite.
However, a particularly interesting case arises when
$\mu_B = 0$, which implies that
\be
\label{starstar} \cosh(\pi s) = 0. \ee
This is solved when $s$ takes
one of a discrete set of imaginary values \be\qquad \quad s = {i\over
2} (2 m + 1) \, , \qquad m \in {\bf  Z}. \ee
Investigation of the
spectrum of open string states associated with each boundary state
$\ket{B_s}$ furthermore reveals that
as one increases $n$ one finds
increasingly tachyonic open-string
modes \cite{Teschner:2000md}.
We will focus on the minimal value $s = {i\over 2}$.
Note that these D-objects
do not have a continuous degree of freedom
corresponding to their space-like position.

To obtain a bit more insight into the structure of the boundary
state $\ket{B_s}$ it is instructive to introduce a boundary state
with fixed {\it length} equal to $\ell$ via \be \label{laplace}
\ket{B_s} = \int {d\ell \over \ell}\, e^{-\ell\sqrt{\mu} \cosh (\pi s)} \,
\ket{W(\ell)} \ee The overlap of this new boundary state
$\ket{W(\ell)}$ with the momentum eigenstate $\ket{v_P}$ takes the
form \be \label{bessel} W(\ell,P) \equiv \vev{v_P|W(\ell)} =
 c \, e^{-i\delta(P)} \, P \, K_{iP}(\sqrt{\mu} \ell) \ee  Interestingly, this
expression satisfies the Schrodinger equation \be \label{wheeler}
\Bigl( -{1 \over 2} {\partial^2\over
\partial \phi_0^2 } + 2 \pi \mu e^{2 \phi_0} - {1\over 2} P^2 \Bigr) W(\ell,P)
= 0 \qquad \qquad \ell = e^{\phi_0} \ee which is often referred to as
the ``mini-superspace Wheeler-De Witt equation'' of the 2-d
Liouville gravity theory, since it takes the form of the zero-mode
truncation of the Liouville equation of motion. It shows that $W(\ell,P)$ can
be thought of as the space-time profile of the tachyon mode
created by the microscopic vertex operator $V_P$ of momentum $P$.
Applying the
Laplace transformation (\ref{laplace}) to the result
(\ref{bessel}) reproduces (\ref{fuzzy}), via the identity \be
\label{geez}
 \int {d\ell \over \ell} \, \, e^{-\ell \sqrt{\mu} \cosh( \pi s)}
K_{i P} ( \sqrt{\mu} \, \ell) = {\cos (\pi P s) \over P \sinh  (\pi
P )} \ee



Another very interesting class of boundary conditions for the
Liouville theory were studied in \cite{Zamolodchikov:2001ah}.
These boundary states have the property that the only state that
propagates in the open string channel is the identity operator.
They are highly localized in the Liouville direction, and do not
have a space-like position. As pointed out in
\cite{Polyakov:2001af}, this implies that their worldline
description reduces to a matrix model. \footnote{A study of their
interpretation as D-objects, and their description in the matrix
model was recently performed in
\cite{emil}\cite{nowwehaveexactlywhatwewant}. }

In 1+1-d string theory, the Liouville theory is supplemented with the
$c=1$ CFT of the time coordinate $X^0$ and we can thus consider boundary states
that describe a D-particle with a rolling open string tachyon on its world-line.
When the Liouville boundary state has $\mu_B = 0$,
the profile of the open-string
mode arises purely from the $X^0$ boundary state;
otherwise, there is a time-independent term in the
profile.  This is our main motivation for
focussing on $\mu_B = 0$.
For later comparison with the $c=1$ matrix model,
we would like to determine the one-point function that expresses the
leading order emission of closed string tachyons from this boundary state.
For concreteness, we will consider the half-brane trajectory (\ref{roll}).
The corresponding boundary state takes the form of the tensor product
 \be \ket{B} =
\ket{B_\lambda} \otimes \ket{B_s}. \ee We are interested in the
overlap of this boundary state with the state
$\ket{v_\omega}\otimes \ket{v_P}$ created by the vertex operator
$V_{\omega,P} = e^{i\omega X^0 + (Q+iP)\varphi}$ of given energy
$\omega$ and Liouville momentum $P$, subject to the mass-shell
condition that $\omega = |P|$. The time-dependent state
$\ket{B_\lambda}$ has a non-zero overlap with energy eigenstate
$\ket{v_\omega}$ -- normalized according to
$\vev{v_\omega|v_{\omega'}} = \pi \delta(\omega-\omega')$) equal
to \be \vev{v_\omega | B_\lambda} = \lambda^{-i \omega} { \pi
\over \sinh \pi \omega}. \ee So the total production amplitude is
(specializing to $s=\pm i/2$) \be \label{production} \qquad \quad
\CA(\omega, P) =  \vev{v_\omega| B_\lambda}\vev{v_P|B_s} = {c\,
\pi^2 e^{-i\delta(P)} \lambda^{-i\omega}\over \sinh( \pi P/2)
\sinh(\pi \omega)},\qquad \ \omega=|P|\, . \ee In a section five
we will reproduce this exact same amplitude by considering the
emission due to a classical rolling eigenvalue in the $c=1$ matrix
model.


\newsubsection{D-particle Gas in a 1-D Box}

We will now proceed to analyze the quantum mechanics of the
non-abelian tachyon mode that lives on the worldline of (a bound
state of) many unstable D-particles. That we can treat the tachyon
mode in isolation, without coupling to other worldvolume fields,
will be justified for unstable D-particles in the bosonic and type
IIB theories in section 7. In the 1+1-d theory, there are indeed
D-branes with this property. \footnote{It has been pointed out in
\cite{Polyakov:2001af}\cite{joerg}, this analysis is most
naturally applied to the branes of \cite{Zamolodchikov:2001ah}.}
We will assume that the effective action of the $T$ mode is as
given in eqn (\ref{eff}).

First, however, let us address an apparent puzzle. Perturbative
study (for small values of the Liouville interaction $\mu
e^{2\phi}$) of the open 1+1-d string spectrum reveals that, like
its closed string cousin, the open string tachyon in 1+1-d string
theory is not really ``tachyonic'' but rather ``massless'': its
perturbative potential starts out flat, with zero second
derivative instead of with a negative one as in eqn
(\ref{approx}). Indeed, in general the presence of open and closed
string tachyonic modes can be seen as a consequence of the
Hagedorn growth of the number massive string states; the 1+1-d
model has no massive on-shell degrees of freedom, and its physical
spectrum is therefore free of tachyonic instabilities. In other
words, in the perturbative regime of small $\mu e^{2\phi}$,
D-particles in 1+1-d string theory would appear to be perfectly
stable! So how can the action (\ref{eff}), with the unstable
potential (\ref{special}), be a correct effective model?

The justification for (\ref{eff}) is that its classical
trajectories reproduce the consistent tachyon profiles
(\ref{bounce}). This reasoning is still perfectly valid in the
1+1-d theory. The implicit assumption leading to the apparent
contradiction is that a D-particle can be independently localized
in the Liouville direction $\varphi$ as well as in $T$. This
assumption, however, is invalidated by the fact that the type of
D-particle we wish to study has only $T$ as its low energy degree
of freedom, and does not have an independent $\varphi$ position.
This suggests that this $\varphi$ position is in fact correlated
with the value of $T$, and it would seem a good guess that small
values of $T$, near the top of the potential $V(T)$, describe a
D-particle in the strongly coupled region near the ``Liouville
wall,'' while large values of $T$ (near the flat region of the
potential $V(T)$) correspond to $\varphi$ values in the weakly
coupled asymptotic region. All this is of course very reminiscent
of the holographic dictionary of the AdS/CFT correspondence, and,
more to the point, of its precursor: the c=1 matrix model.

The $c=1$ matrix model describes 1+1-d string theory in terms of the
quantum mechanics of a single large $N$ matrix. It has long been recognized,
ever since the D-brane string revolution, that the eigenvalues
of this large $N$ matrix have a likely interpretation as
the positions
of D-particles of a suitable type. Given the above discussion,
it is natural to suspect that the appropriate D-particle is that of
the 1+1-d theory itself. The effective lagrangian describing $N$
such D-particles is
\be
\label{smatrix}
S(\TTT) = - \int\! \! dt \, {\rm{Tr}}\, \Bigl(\,
V(\TTT) 
\sqrt{1-(D_t{\TTT})^2}\, . \Bigr)\ee
Here $\TTT$ is a an $N\times N$ hermitian matrix and
$D_t=\partial_t + [A_t, \, \cdot\,]$,
with $A_t$ a $U(N)$ gauge ``field''.
Choosing $A_t = 0$ gauge, its only role is to
impose the Gauss Law which projects onto
singlet states.
Given the form (\ref{special})-(\ref{approx}) of $V(T)$, it is
clear that the quantum mechanics of this model has the same
large $N$ behavior as the $c=1$ matrix model. In this section we
will make this correspondence explicit, while recollecting
some relevant facts about the $c=1$ matrix quantum mechanics.

Following the standard $c=1$ routine
\cite{longlist}
we write
\be
\qquad \TTT = \Omega^\dagger T \, \Omega \, , \qquad \quad
\qquad \ T = {\rm diag}(\TT_{\! 1}, \TT_{\! 2}, \ldots\, \TT_{\! N}).
\ee
Here the eigenvalues $T_i$ parametrize the positions of the D-particles,
while the $U(N)$ matrix $\Omega$ represents a pure gauge degree of freedom.
Thus we may set $\Omega=1$. Taking into account the proper Jacobian, we
can write the quantum mechanical wave-function of $\TTT$ as
\be
\label{redux}
\Psi(\TTT) = \Delta(T) \Psi(T)\,, \qquad \qquad
\Delta(T) = \prod_{i<j}(\TT_{\! i}-\TT_{\!\! j}).
\ee
Since $\Delta(T)$, the Vandermonde determinant, is completely
anti-symmetric under interchange of the eigenvalues $T_i$, the reduced
wave-function $\Psi(T)$ is completely anti-symmetric as well. The Hamiltonian
derived from (\ref{smatrix}) reduces to a sum of decoupled single particle
Hamiltonians
\newcommand{\muf}{\mu_{{}_{\! F}}}
\newcommand{\hh}{{\mbox{\large $h$}}}
\newcommand{\ppi}{{\mbox{\Large $\pi$}}}
\be
\label{ham}
H = \sum_i \hh_i \, ,\qquad \quad \hh_i = \sqrt{\, \ppi_i^2 + V(\TT_i)^2}
\ee
with $\pi_i = i{\partial \over \partial T_i}$. The $N$ D-particle matrix model
thus reduces to a decoupled system of $N$ fermions in one dimension, described
by the relativistic Hamiltonian (\ref{ham}).

{} From the form of the potential $V(T)$ we see that the single particle
Hamiltonian $h$ has a continuous spectrum of plane wave eigenstates.
To resolve this continuum into a discrete spectrum, we will put the
system in a 1-d box, by imposing reflecting boundary conditions at
at some large (negative) value $T=T_0$. We will choose this cut-off
location at
\be
\label{tnot}
T_0 =  2\log g_s.
\ee
Eventually we will remove this IR cut-off taking the limit $g_s \to 0$.

The ground state of the system is obtained by filling the lowest $N$
eigenvalues of $h$. We will assume that $g_s$, $N$ and $T_0$ are tuned
to that the fermi level defined by the $N$-th energy eigenvalue
\be
\muf \equiv E_N
\ee
lies just below the top of the potential barrier
\be
\qquad
\muf {{\mbox{\raisebox{-5pt}{$<$}}}\atop
{\mbox{\raisebox{3pt}{$\sim$}}}} \, \mu_c \, , \qquad \quad \ \mu_c \equiv
 V(0) = {1\over g_s}.
\ee
Using the WKB approximation, and ignoring for now the quantum mechanical tunneling
through the potential barrier, we can find the fermi level $\muf$ for given $N$
via the Bohr-Sommerfeld condition
\be
\label{nnot}
\qquad \qquad 2\pi N = \oint
\! d\TT  \, \ppi_{{}_{\! N}}(\TT) \, , \qquad \qquad \ppi_{{}_{\!
N}} \! =\! \sqrt{\, \muf^2 \! -\! V(\TT)^2}\, , \ee where the
integral is around the closed trajectory with energy $\muf$. Thus
\be N = {1\over \pi} \int\limits_{\TT_{\! 0}}^{\ \TT_{\! F}} \!\!
d\TT
\sqrt{\muf^2 - V(T)^2} 
\ee
with $\TT_{{}_{F}}$ the turning point at which $V(\TT_{{}_{F}}) = \muf$.

An interesting quantity is the spectral density $\rho(\muf) =
{\partial N \over \partial \muf}$, which reveals a logarithmic
divergence in the limit that $\muf$ approaches $\mu_c$: \be
\label{rho} \rho(\muf)  =   {1\over \pi} \int\limits_{\TT_{\!
0}}^{\ \TT_{\! F}}
\! {\muf dT \over \sqrt{\muf^2 - 
V(\TT)^2}} \; 
\, \simeq\,  -{2\over \pi}  \log \mu,
\ee
with $\mu \equiv \mu_c\! -\!\muf$.
Since we have set the cutoff on the eigenvalue space
at $ \TT_0 = 2 \log g_s$, (\ref{rho})
gives the expected linear growth of the spectral density with the 1-d
volume.
Further, the $\log(\mu)$ divergence
reflects the
fact that the time that a particle spends near the top of the potential
barrier diverges logarithmically for small $\mu$. Correspondingly, its
wave-function is sharply peaked near the turn-around point.


We thus reach the -- with hindsight not entirely surprising -- conclusion
that the gas of D-particles in 1+1-d string theory, with chemical
potential $\muf$ close to $\mu_c$,
displays the exact same universal behavior as the $c=1$ matrix
model. This correspondence becomes exact in the double scaling limit:
\be
g_s\to 0, \ \ N \to \infty, \ \ \ \mu= {\rm fixed}.
\ee
Essential for this correspondence is that (i) the D-particles
have only one matrix coordinate $T$, and (ii) the effective potential
$V(T)$ has a single maximum of the upside-down harmonic form (\ref{approx}).

\figuur{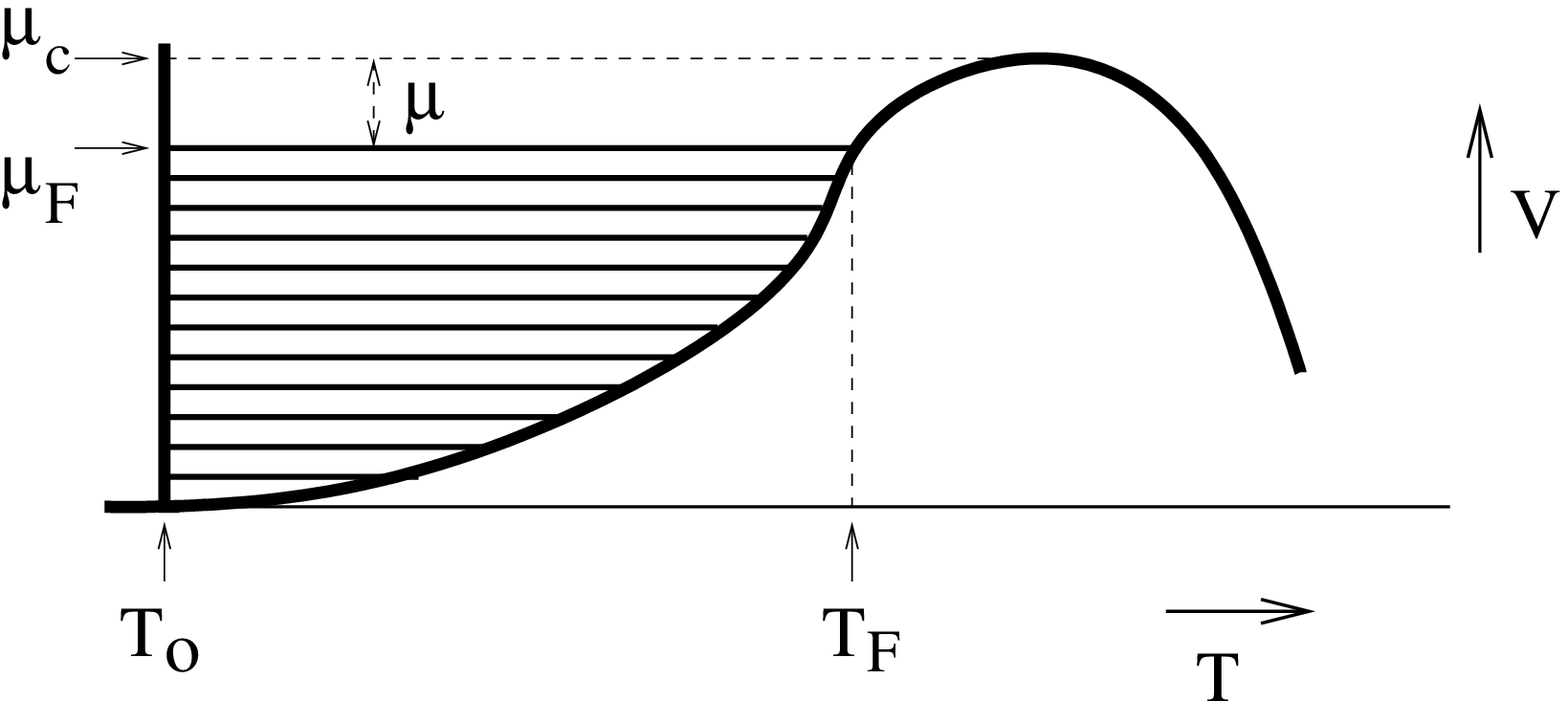}{10cm}{Fig 1. The tachyon potential $V(T)$ and the
Fermi sea of filled energy eigenstates. The difference $\mu$
between the Fermi level $\muf$ and the top of the potential
specifies the effective string coupling of the dual 1+1-d string
theory.}

Much work has been done to extract the scattering amplitudes
from the $c=1$ matrix model and to compare them with
the corresponding string theory computations and expectations.
While the duality has passed many checks,
several important puzzles remain. We list of some of the
main results and open questions:

\noindent
(i) The target space dynamics of the matrix model can be formulated
in terms of a collective field theory \cite{dj}, in which the eigenvalue
density $\rho$ appears as a second quantized 1+1-d field. This collective
field is expected to be related via a field redefinition to the
closed string tachyon of the non-critical string theory.
(In connection with this approach, it is perhaps interesting to note that
the relativistic form of the eigenvalue Hamiltonian (\ref{ham}) allows
us to introduce a second quantized fermionic formulation based on
the rather standard looking action
\be
S(\Psi)= \int\!\! d^2 \XX \,
\Bigl(\, \overline{\Psi} \,
\Bigl(\gamma^\mu {\mbox{\Large $\partial \ \over \partial X^\mu$}}
-  V(\XX^1)\, \Bigr) \Psi \,  +
\muf (\Psi^\dagger\Psi - N)\Bigr)
\ee
where $\Psi$ denotes a 1+1-d Majorana fermion. Collective
field theory essentially arises via bosonization of the fermion
$\Psi$.)

\noindent
(ii) Exact expressions for string scattering amplitudes have been
found to all orders in the effective string coupling constant
\be
g_{\eff} = {1/\mu}.
\ee
These expressions have been checked against tree-level string
theory calculations.

\noindent (iii) While the scattering amplitudes are
perturbatively unitary, the matrix model reveals an inherent
non-perturbative instability leading to unitarity violations of
order $e^{-2\pi\mu}$ \cite{Shenker:1990uf}. 
The instability is caused by tunneling
events of eigenvalues through the potential barrier. The tunneling
amplitude \be \label{gone}
\exp\Bigl({-\!\!\int\limits_{{}^{-T_F}}^{{}^{T_F}} \!\!\! dT
\sqrt{\, \mbox{\raisebox{0pt}{$V(T)^2\! -\muf^2$}}}}\, \Bigr)\;
\simeq \;  \exp({-2\pi \mu}) \ee reveals the characteristic
$e^{-O(1/g_{\eff})}$ behavior of a D-brane process.
Historically,
this result of course preceded \cite{Polchinski:fq}
and in fact precipitated the later
discovery of the role D-branes.
With hindsight, the identification
of the eigenvalue tunneling event with the D-instanton
leads directly, via
an application of Sen's descent relations \cite{Sen:99},
to the equivalence between
eigenvalues and D-particles.

The presence of these unitarity violations turned out to be a
severe problem for the $c=1$ model, and it is as yet unclear
whether these can be remedied without violating target space
locality and/or causality. (For two divergent opinions on this
question see \cite{Polchinski:1994jp} and \cite{Dhar:1995gw}.)
 This of course seriously dampened the
enthusiasm about the $c=1$ matrix model as a potential
non-perturbative formalism for 1+1-d string theory: it is now
considered to be an incomplete model, without any consistent
and/or acceptable non-perturbative definition. Given its key role
in the inception of the D-brane era, however, the model deserves a
better fate than that. Our proposed identification of the matrix
eigenvalues with the tachyon mode of D-particles hopefully
represents a first step towards its rehabilitation. We will come
back to this issue in section 7, where we propose a possible way
for embedding the $c=1$ matrix model into IIB string theory.

\newsubsection{Closed-String Radiation from a Rolling Probe-Eigenvalue}

The surfaces dual to matrix-model diagrams have boundaries if the
matrix model is coupled to variables in the fundamental
representation of the matrix quantum mechanics gauge group
(This fact has been emphasized by
{\it e.g.} \cite{Yang:hs}\cite{Minahan:1992bz}\cite{Kazakov:1991pt}).
Exactly such
variables arise when a classical probe eigenvalue is introduced
into the system, extending the rank of the $N\times N$ matrix
$\TTT$ by one extra row and column \be \TTT_{{}_{N+1 \times \,
N+1}}= \left(\matrix{ z & v_1 & v_2 & \dots \cr v_1^* & T_{11} &
T_{21} & \dots \cr v_2^* & T_{12} & T_{22} &  \dots \cr \dots &
\dots & \dots & \dots
} \right). \ee For fixed extra eigenvalue $z$, the variables $v_i$
have single line propagators that delineate a boundary of the
string world-sheet, with a Dirichlet boundary condition located at
$z$. As a quantitative test of this identification between
eigenvalues of the $c=1$ matrix and D-particles in 2D string
theory, we will compute the closed string radiation produced by a
classical rolling extra eigenvalue in first-order time-dependent
perturbation theory. For the probe trajectory we will take \be
\label{rolll} z(t) = \lambda\,  e^t. \ee The corresponding
perturbation hamiltonian arises as follows.

\def\rb{{\mbox{\large$)$}}}
\def\lb{{\mbox{\large$($}}}

Suppose we start from the  $N\!+\! 1 \times N\!+\! 1$ matrix model. We can
write its wave-function in terms of its eigenvalues and after
splitting off the Vandermonde determinant obtain the free
Hamiltonian $H_{N+1} = H_0 + H_{probe}$. Here $H_0$ is the free
hamiltonian of the $N\times N$ model and $H_{probe}$ that of the
extra eigenvalue. We want to make the probe follow a classical
trajectory. We should therefore consider a wave-function of the
$N$ quantum mechanical eigenvalues only, and rather than factoring
out the complete $N\! +\! 1\times N\! +\! 1$ Vandermonde determinant, we
should split off the $N\times N$ determinant only.
As a result, the Hamiltonian $H_{N+1}$ does not reduce to the free
Hamiltonian $H_0$ of $N$ non-interacting fermions, but takes the
form \be \label{conjug} H(t) = e^{-W(z(t))} H_0 \, e^{W(z(t))}\, ,
\ee with \be \tilde W(z) = {\rm Tr} \log (z - \TTT ) . \ee
Note that this object may be thought of
as the gauge parameter inducing a background worldvolume gauge field
created by the probe.
Since
the large $N$ theory is semi-classical, we can simplify
(\ref{conjug}) to \be H = H_0 + H_1\, , \qquad \qquad H_1 = [ H_0,
\tilde W\lb z(t)\rb]. \ee

We want to compute the transition amplitude from an initial state
$\ket{\muf}$, in which the Fermi sea of the $N$ eigenvalues is
calm and quiet, to a final state $\ket{\muf + \omega}$ with a
single excitation with energy $\omega$.
In first-order time-dependent perturbation theory,
this amplitude takes the form \be
\label{ohm}
\CA(\omega) = \int \!\! dt\, \bra{\muf+ \omega }\,
 [ H_0, \tilde W\lb z(t)\rb]\,
\ket{\muf}  = \omega \int \!\! dt\, \bra{\muf + \omega } \, \tilde
W\lb z(t)\rb \ket{\muf} \ee 
The matrix operator $W(z)$ is related
via a Laplace transform to the ``macroscopic loop operator''
$W(\ell)$ that creates a boundary to the string world sheet with
total boundary length $\ell$, 
with a Dirichlet condition\footnote{
We thank N. Seiberg for a discussion on this point.}
at time $t$:
\be \tilde W(z, t) =
\int\limits_0^\infty \!{d\ell \over \ell}\, e^{- z \ell}
~W(\ell, t)\, \qquad \qquad W(\ell, t) \equiv {\rm Tr} \, e^{\ell \,
\TTT(t) }. \ee The relevant matrix elements of the macroscopic loop
operators $W(\ell)$ have been evaluated in 
\cite{Moore:1991ir}\cite{Kazakov:1991pt}, with the result \be \label{loop}
\vev{\muf+\omega| W(\ell,t) |\muf } = e^{-i\delta(\omega)}
~e^{i \omega t}~
K_{i\omega} (\sqrt \mu \, \ell ) 
\ee 
with $\mu = \mu_c-\muf$ as
before, and where we have included the phase
$e^{-i\delta(\omega)}$ as defined in equation (\ref{leg}). This phase
is not produced by the matrix model itself, but appears as a
separate wave-function renormalization, the so-called leg-pole
factor, that relates the matrix model states to the continuum
closed string tachyon modes (see for example Eqn. (7.27) of
\cite{Klebanov:1991qa}).

Combining (\ref{ohm}) with (\ref{loop}) and the
integral formula (\ref{geez}) gives \be
\label{needsaname}
 \CA(\omega) =
\int \!\! dt\, e^{-i\delta(\omega)}~e^{i\omega t}~ 
{ \cos \lb\pi \omega \, s(t)\rb \over \sinh (\pi \omega)}
\ee where $s(t)$ parametrizes the probe trajectory via
\be z(t) = \cosh\lb \pi s(t)\rb .\ee 
At this point we can already point out a 
striking correspondence with the 
boundary states $\ket{B_s}$.
Namely, we can write $\CA(\omega)$ as \cite{joerg}
\be
\label{joergformula}
\CA(\omega) = \int dt ~ \bigl[
\vev{\omega | B^D_t} 
~\vev{ P | B_{s(t) }} \bigr]_{\omega = P}
\ee
where $ \ket{ B_t^D} $ is a 
Dirichlet boundary state at time $t$.
Now for the rolling
trajectory (\ref{rolll}) we have
 \be \label{probeelement}
\int \!\! dt\, \cos (\pi \omega s(t))~e^{i\omega t}
 = \int \!\!
d s \, \rho(s) \, \cos (\pi \omega s)
~e^{i\omega t(s)} \ee with \be \rho(s) =
{dt\over ds} =  {\pi \over 1 + e^{2\pi s}} - {\pi \over 1 +
e^{-2\pi s}}. \ee We can view this quantity $\rho$ as the
semi-classical expression of the quantum mechanical probability
density $|\psi(s)|^2$ of the probe eigenvalue. It is interesting
to note that this expression has poles exactly at the special
values of the Liouville boundary state parameter $s$ discussed
earlier \be \qquad \ \ \ s ={i\over 2} (2m+1) \qquad  \ \ m \in
{\bf Z}. \ee These values of $s$ correspond to the location $z=0$
at the top of the inverted harmonic potential. The physical origin
of the poles in $\rho(s)$ is of course that the critical
trajectory (\ref{rolll}) has zero velocity at this location, while
the periodic recurrence of the poles reflects the periodic orbits
of the right-side up harmonic oscillator.


To perform the integral
(\ref{probeelement}), we can write $\cos(\pi\omega s) ={1\over 2}( e^{i\pi \omega s}+e^{-i\pi \omega s})$ and in
each of the two terms we can 
attempt to close the contour in the upper and lower
half complex $s$-plane respectively.  
The contributions from the 
sum over residues gives \be
\sum {\rm Res}~\rho(s) \, \cos (\pi \omega s )\,
= {\pi \over \sinh (\pi \omega/2)}.
\ee
This contributes a term in production amplitude
\be
\label{repro}
\CA_{\rm Res}(\omega) = {\pi \, e^{-i
\delta(\omega)} \lambda^{i \omega t}
\over \sinh (\pi \omega/2) \sinh(\pi \omega)}, 
\ee
which 
reproduces 
the continuum production amplitude (\ref{production}) 
of the rolling tachyon 
of the $\ket{B_{s=i/2}}$ state.
This correspondence, 
but more accurately the match 
between (\ref{joergformula}) and 
(\ref{needsaname}), 
supports
our interpretation of the extra
eigenvalue as an unstable D-particle of the 1+1-d string theory.
The expression (\ref{repro}) omits, however, 
a long-time divergence arising from the countour at 
infinity.
A precise correspondence involves a somewhat different 
interpretation of the matrix model computation 
in terms of boundary states, and is discussed further 
in \cite{joerg}.

It is in principle straightforward to compute the higher order corrections to the
leading order result (\ref{repro}). To find a sensible total production rate, however,
one would need to treat the probe quantum mechanically, so that the recoil is included
and total energy is preserved. It is evident from the matrix description that
all the initial energy (of order $1/\mu$) of the D-particle will be emitted in
the form of closed string tachyon radiation.

\newsubsection{Large $N$ RG Formulation of the Duality}

The matrix model
description in section 4 of $N$ D-particles in 1+1-d string theory reveals a
universal behavior that depends only on one parameter: the
effective string coupling $\mu$ (which e.g. sets both the size of the
tunneling amplitude (\ref{gone})) as well as of the eigenvalue density
$\rho(\muf)$ as given in (\ref{rho})). The model itself, however,
has two parameters: the number of D-particles $N$ (all assumed
to be in their lowest possible energy eigenstate), the (bare)
string coupling $g_s$.
We are therefore led to conclude that there should be an
equivalence between 1+1-d string theories with different values of
$N$, $g_s$, but with the same values for $\mu$:
\be \label{rgdual} \qquad \quad (\, g_s, N\,
) \; \cong (\, \tilde{g}_s, \tilde{N}\, ) \qquad
{\rm if} \qquad
\mu = \tilde{\mu}
\ee This is our new proposed duality relation, and can be thought
of as a large $N$ renormalization group transformation that leaves
1+1-d string theory invariant.

\newcommand{\nulll}{{{}_{0}}}

A special case of this duality relation is the equivalence of a
string theory with $N$ D-particles and coupling $g_s$ and the
string without D-particles and string coupling $g_{\eff}=1/\mu$.
This equivalence reduces in the large $N$ limit to the usual $c=1$
matrix model duality. As long as $N$ is finite, the string
coupling $g_s$ is finite as well, so that the
D-particles are not decoupled: they must be described as embedded and
interacting with the surrounding closed string theory. The finite $N$ theory
thus gives an interesting hybrid description
of the effective closed string theory, in which the effective world-sheets are
partially made up from
open string diagrams and partially from
closed string diagrams. Only upon taking the double scaling
limit $N\to \infty$, $g_s \to 0$ with $\mu$ fixed, can we treat the
effective D-particle action (\ref{smatrix}) as a strictly
decoupled system. Thus our proposed duality gives a new physical
explanation of the origin of the double scaling limit.

Another interesting case is $N =1$ and $\tilde{N}=0$. It entails
that adding a single D-particle in its lowest energy eigenstate,
can be absorbed via a small adjustment of the string coupling.
This statement appears to be closely related to the
aforementioned result that the critical time-dependent tachyon
profile (\ref{bounce}) with $\lambda={1\over 2}$ is equivalent to
a source free wave-like solution of the closed string background
fields. Indeed it is natural to interpret the process of adding an
extra D-particle in its lowest energy eigenstate as the quantum
description of the classical $\lambda ={1\over 2}$ tachyon bounce.
Since we have placed the system in a box by adding a reflecting
boundary condition at $\TT_0$, the minimal energy bounce will
instead take the form of a minimal energy standing wave. This
predicts that, on the closed string side of the duality, the
$\lambda={1\over 2}$ wave will also relax to a minimal energy
standing wave, which according to our proposal is just a small
static shift of the closed string background fields.

To further substantiate this physical equivalence, reconsider the
above emission calculation for the special probe trajectory \be
\label{bounce2} \qquad z(t) = 2\lambda \sqrt{\mu} \cosh t \qquad
\qquad \lambda={1\over2}. \ee This trajectory precisely follows
the surface of Fermi sea. In terms of the variable $s(t)$ it is
described by $\pi s(t)= t$. The corresponding closed string
production amplitude is therefore proportional to $\int ds
\cos(\pi \omega s)$ and thus vanishes for all non-zero $\omega$.
It thus corresponds to a static shift in the closed string
background, as advocated. Sub-critical trajectories with
$\lambda>{1\over 2}$ are obviously Pauli excluded, while
super-critical trajectories with $\lambda<{1\over 2}$ do generate
non-trivial emission amplitudes. Eventually, all super-critical
trajectories decay to the minimal energy one; in the target-space
field theory, the only permanent remnant of the presence of the
extra D-particle is a small adjustment of the tachyon background
proportional to the associated small shift in the Fermi sea.
In the Appendix we show how this shift is calculated from the continuum
boundary state description of section 3.

It is instructive to consider the dual equivalence (\ref{rgdual})
with $N ={\tilde N} +1$,
which is the smallest renormalization group step. To obtain the
explicit form of the corresponding
background shift, it turns out to be a bit more practical to first
view $N$ as a
function of the coupling $g_s$, whose form is determined via the
condition that $\mu$ is held fixed.  This
condition results in a differential equation, which is easily found
from (\ref{nnot}) and (\ref{tnot}).

Let us summarize. We have formulated a new open/closed string duality
relation (\ref{rgdual}) between 1+1-d string theory backgrounds
with different numbers of D-particles. This duality provides a new
interpretation and physical foundation of the $c=1$ matrix model,
which hopefully will help in putting the string/matrix model
duality on a somewhat firmer footing.
A problem that remains, however, is that the 1+1-d string theory
is non-perturbatively unstable, although we now have a more direct
interpretation of this instability as the decay of a D-particle to
the ``wrong side'' of its tachyon barrier, where the potential is
unbounded from below. Clearly, it would be of interest to find a
consistent completion of 1+1-d string theory, for example by
embedding it in a larger self-consistent framework. In the
following section we will give a concrete proposal in this
direction in terms of IIB superstring theory.

\newsubsection{c=1 Matrix Model from IIB String Theory}

In this section we will consider a dense gas of unstable
D-particles in IIB superstring theory, and argue that, in a
suitable decoupling limit, its description reduces to the $c=1$
matrix model.

In comparison to our discussion of D-particles in the
1+1-dimensional bosonic string theory, there are several new
ingredients that we need to take into account. First, unstable
D-particles in IIB string theory have, besides the open string
tachyon mode, also other light degrees of freedom, namely their
positions $X^0$.\footnote{The worldline theory 
of an unstable D-particle 
in type IIB string theory 
also contains 32 worldline fermions,
whose presence will be ignored in the following 
discussion.}
The non-abelian worldline action of $N$ particles
therefore has a more complicated form \cite{Garousi:2000tr},
\cite{Kluson:2000iy}, \cite{Bergshoeff:2000dq}
\be \label{sdbi}
S_{DBI} = \int \!\! dt {\rm Tr}\left( V(\TTT) \sqrt{\Bigl((1 -
(D_t\TTT)^2 + (D_t X^i)^2\Bigr) \det\lb\delta_{ij}+[X_i,X_j]\rb} +
f(\TTT) [X_i,\TTT]^2 + \ldots\right) \ee Our goal is to show that,
in a suitable high density limit, the tachyon mode $T$ becomes
much lighter than the $X^i$'s, so that in this limit the model in
fact reduces to the $c=1$ matrix quantum mechanics.

Besides the Born-Infeld action (\ref{sdbi}), the IIB D-particle
world-line action also involves a Chern-Simons term that describes
its coupling to the RR scalar, the IIB axion field $C$. To write
this term, let us momentarily ignore the other D-particle degrees of freedom,
and concentrate on the tachyonic mode only.  The complete action,
including the Chern-Simons
coupling \cite{Sen:2002nu}\cite{Okuyama:2003wm}\cite{Sen:2003tm},
then reduces to (here the \ ${}'$ \ denotes derivative with
respect to $T$) \ba \label{sred} S \is  S_{DBI} + S_{CS} \\[3mm]
\is \int \!\! dt\, {\rm Tr}\Bigl( V(\TTT) \sqrt{1-(D_t
\TTT)^2}\Bigl)  \; +\; \int \!\! dt\, {\rm Tr}\Bigl( C\,
W^\prime(\TTT) D_t \TTT\Bigr), \nonumber \ea where
\footnote{Notice that $V(\TTT)$ is identical to its bosonic cousin
(\ref{special}) up to a rescaling of $T$ by a factor of
$1/\sqrt{2}$. To understand this factor, recall that the intercept
in the fermionic string is half that of the bosonic string. In
boundary CFT language: to turn on an open string tachyon profile
$T(X^0)$, one needs to introduce a boundary interaction of the
form $$S_{bdy} = \int \!\! d\xi \, \psi^0(\xi) T^\prime(X^0(\xi))
\otimes \sigma_1$$ with $\sigma_1$ a Chan-Paton index
\cite{Sen:99}. Conformal invariance thus requires that $T(X^0)$
has scale dimension $1/2$.} \be \label{ww} W^\prime(T) = g_s V(T)
= {1\over \cosh(T/\sqrt{2})} \ee As was recently shown in
\cite{kutasov:2003er} and \cite{Okuyama:2003wm}, this form is
completely fixed by the requirement that the known consistent open
string tachyon profiles $T(X^0)$ solve the equation of motion of
(\ref{sred}). Note further that the relation (\ref{ww}) between
$W(T)$ and $V(T)$ ensures that a D-instanton, which is known to
correspond to a trajectory $T(t)$ that runs from the minimum at
$T=-\infty$ to the other minimum at $T=+\infty$, carries the
correct unit of RR-charge.

Motivated by the preceding discussion of 1+1-d string theory, let
us consider a dense gas of $M$ unstable D-particles inside of some
finite volume $V_9$. We wish to study this system in its lowest
possible energy state. As we have learned, this means that the
tachyon mode $T$ on each D-particle must follow the minimal bounce
trajectory (\ref{bounce}) with $\lambda={1\over 2}$. This minimal
trajectory is called an sD-brane in \cite{stherm}, where it was
shown that it has the characteristic property that it creates half
a unit of flux for the time-derivative of the axion $C$. $M$
sD-branes inside of a 9-volume $V_9$ thus produce a flux \be
\label{cflux}
\int_{V_9}\!\!\!\! {\mbox \large *} \, d C =  {1\over 2} \, M .
\ee
Positive and negative $M$ correspond to sD-branes with
positive and negative $T(t) = \pm {1\over 2} \cosh t$.
If we assume that the particles are evenly distributed, we
conclude that every particle, via the flux produced by all the other
particles, is immersed in a uniform field \be \label{cnot} \qquad
\qquad \qquad \partial_0 C = {1\over 2} \,  \nu\, , \qquad \qquad
\nu = {M\over V_9}. \ee

How should we incorporate the presence of this background flux
into the effective action of the $M$ D-particles? Here we need to
be a bit careful. It is tempting to conclude that we need to include
a non-zero $C$ in the
Chern-Simons term:
\footnote{Here we are dropping a boundary term in
the partial integration.} \be \label{cs2} S_{CS} = {1\over 2} \int \!\! dt\,
\nu \, t \, {\rm Tr} \Bigl( W '(\TTT) D_t \TTT \Bigr) = - {1\over 2} \int \!\!
dt\,\nu \,{\rm Tr}\, W(\TTT)\, . \ee However, since the matrix variables
of the non-abelian DBI action include the open string states that
stretch between the particles, via the usual open/closed string
equivalence it already includes the effect of closed string exchange
between the particles! We would therefore be double-counting if we
add the CS-term as well.

Now let us instead consider a gas of $N+M$ D-particles inside a
small volume $V_9$, with $M$ large compared to $N$. Let us choose
a localized cluster of $N$ of these particles, and consider them
as moving in the closed string background geometry produced by the
surrounding gas of $M$ D-particles. The non-abelian DBI action of
the $N$ particles now includes the CS-term (\ref{cs2}).
The total tachyon effective potential therefore reads \be
V_{\eff}(\TT) = V(\TT) -  {\nu\over 2} \, W(\TT) \ee which for the
explicit form of potentials (\ref{ww}) reads: \be V_{\eff}(\TT) =
{1\over g_s \cosh(\TT/\sqrt{2})} \, - \, \nu \, \sqrt{2}\, {\rm
arctan}\Bigl(\sinh(\TT \sqrt{2}) \Bigr) \ee The second term will
only become important at very high densities, when $\nu$ is order
$1/g_s$. In the following we will assume that there is no
fundamental obstruction against preparing the system at such a
high density.\footnote{In any case it is clear that, as a
consequence of the scaling limit we are about to take, there is no
obstruction from gravitational collapse. The gravitational radius
of a region containing $M$ critical D-particles is $\ell_s (
\lambda_{{\rm 't~Hooft}} )^{1/7}$, where the 't Hooft coupling is
$\lambda_{{\rm 't~Hooft}}  = g_s M$.  In our double scaling limit,
$\lambda_{{\rm 't~Hooft}}  \to 0$; the string coupling is weaker
than in a generic 't Hooft limit. The gravitational curvature is
of order the string scale, which in the $g_s \to 0$ limit is much
below the Planck scale.}

 \figuur{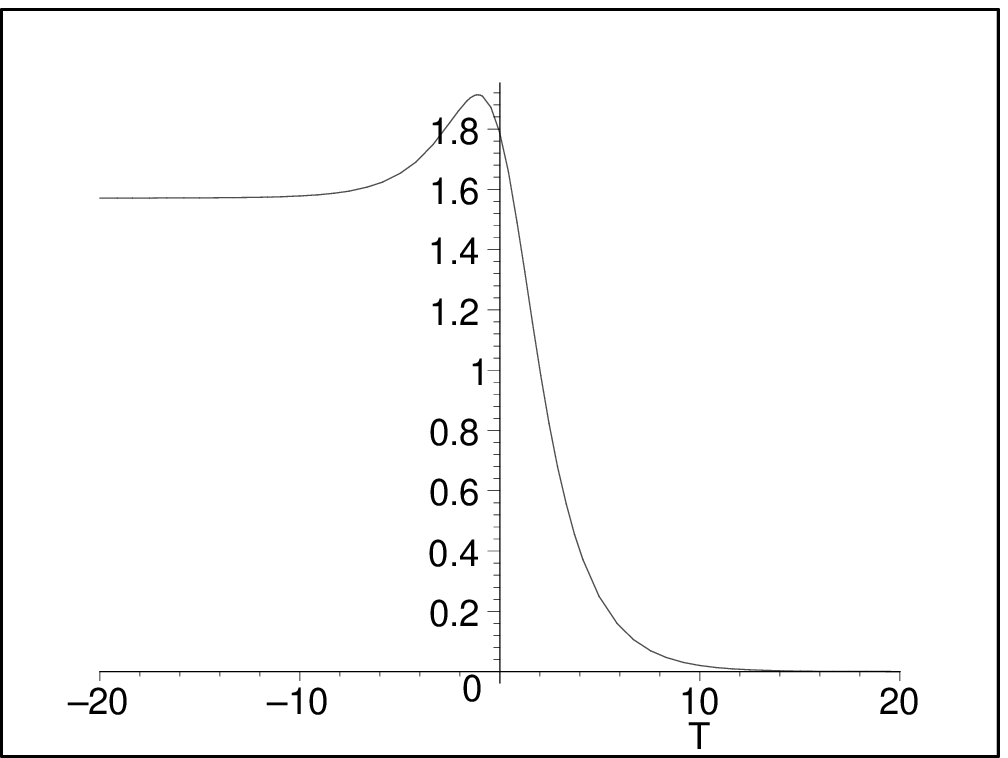}{6.5cm}{Fig
2. Typical form of the effective tachyon potential $V_{\eff}(\TT)$
of unstable D-particles in an axion background with $\partial_0 C
= \nu$ with $\nu$ a constant of order $1/g_s$.}

The typical form of the effective potential is drawn in fig 2. The
effect of the $\nu$ term is to raise the left minimum and lower
the right minimum of $V_{\eff}(\TT)$. In light of our earlier discussion, it seems
natural to  interpret this extra term as the rise in the Fermi sea of the tachyon
matrix eigenvalues due to the presence of the dense gas of D-particles.
In any case, this term has the consequence that tachyon modes that
approach from the left, the potential barrier can be made arbitrarily small.

Let us make this explicit. The effective potential
$V_{\eff}(\TT)$ has a maximum at
\be
\TT_c
\! =\! \sqrt{2}\, {\rm
arctanh}(g_s \nu \sqrt{2}),
\ee
which exists as long as $\nu$ is less then a critical value
\be
\nu_c
= (\sqrt{2}g_s)^{-1}.
\ee
In the limit where $\nu$ approaches $\nu_c$
\be \epsilon \equiv 1- {\nu\over \nu_c}
<\!\!< 1 \, , \ee the effective single particle Hamiltonian
(obtained after reducing of the matrix quantum mechanics to that
of the eigenvalues) near the maximum takes the following form\
\be \label{heff} \hh(\TT_i) \simeq \alpha\;  
\Bigl( {1\over \beta^2} \, 
e^{\hat{T_i}}\, \hat{\ppi}_i^2 \, + \, e^{-\hat{T}}\! -{1\over
3} e^{-3\hat{T}}
\Bigr) \ee
and \be \alpha =\,{\sqrt{2} \epsilon^{3/2}\over g_s}  , \qquad
\qquad \beta = {2 \epsilon\over g_s}. \ee
Here we redefined
$\hat{\TT}_i = {1\over \sqrt{2}} (\TT_i -\TT_c)$, and
$\hat{\ppi}_i = \sqrt{2}\ppi_i$,  so that the new effective
potential has its maximum at $\hat{T}=0$. [\, We have writen
the Hamiltonian in the non-relativistic form, which is justified
as long as the energy is small compared to $V(\TT)$. In the new
variables, this means that we must restrict to the region in which
$e^{\tilde{T}}$ is small compared to $\beta^2/\alpha$. As in the
previous section, we will put a cut-off $\TT \leq \TT_0$ on the
tachyon mode, with $\TT_0$ small enough to satisfy this condition.
In the following, we are going to take the limit $\alpha \to 0$
and $\beta \to \infty$, so this restriction will in fact become
irrelevant.]

\newcommand{\tmuf}{\tilde{\mu}_{{}_{F}}}

{}From the form (\ref{heff}) of the effective Hamiltonian we
deduce that, in this regime with $\alpha$ very small and $\beta$
very large, the spectrum of $\hh$ will contain a large number (of
order $\beta$) of very small eigenvalues (less than $\alpha$). We
would like to use this fact to determine a precise limit in which
the $\TT$-dynamics decouples from all the other degrees of
freedom, and in particular from the space-time motion $\XX_m(t)$
of the D-particles.

The D-particle motion is governed by the matrix action (omitting
factors of order 1) \be S(\XX) = {\epsilon^{1/2} \over g_s}
\int\!\! dt \, {\rm Tr}\Bigl( e^{-\hat{\bf T}} \bigl\{ (D_t
\XX_m)^2 + [\XX_m,\hat{\TTT}]^2 +
[\XX_m,\XX_n]^2\bigr\}\Bigr) \ee The corresponding Hamiltonian
reads \be \hh(\XX) = \alpha 
{\rm Tr} \Bigl(\, {1\over \beta^2}\, e^{\hat \TTT}
P_m^2 + {1\over \epsilon} \Bigl\{ [\XX_m,\hat{\TTT}]^2 +
[\XX_m,\XX_n]^2\Bigr\}\Bigr) \ee We would like to obtain an
estimate of the ground state energy and of the energy gap of this
Hamiltonian.

The classical potential in ${\hh}(\XX)$ has flat directions
$[\XX_m,\hat{\TTT}] = [\XX_m,\XX_n]=0$. These flat
directions are well-known to be lifted by quantum corrections.
(Recall that the world-line theory of unstable D-particles is not
supersymmetric, so there is no cancellation of bosonic vacuum
energies.) 
If we set $\hat{\TTT} =0$, a simple scaling
argument then shows that the remaining hamiltonian
$\tilde{\hh}(X)$ has a non-zero ground state energy and energy gap
proportional to \be \label{de}
{{\mbox{\small $\Delta {E}$}}}
\simeq {\cal O}\Bigl( \, {\alpha \over \beta^{4/3} \epsilon^{1/3}}\Bigr)
= {\cal O} \Bigl(\, {\alpha\, g_s^{4/3} \over \epsilon^{5/3}}\, \Bigr). \ee
Now, in order to achieve the decoupling of the tachyon mode from
the dynamics of the $\XX$-degrees of freedom, we would like this
energy to be much larger than the effective potential of the single
eigenvalue hamiltonian $h(\hat{\TT}_i)$, which is of order $\alpha$.

We are now in a position to give a precise characterization of the
decoupling limit. We are going to send \be \label{one} N \to
\infty\, , \qquad \qquad g_s \to 0\, , \qquad \qquad \epsilon \to
0. \ee We keep $\beta N$ fixed \be \label{two} {g_s N \over
\epsilon} \qquad {\rm fixed}. \ee Further, we want to make sure
that the energy scale {\small $\Delta {E}$} in
(\ref{de}) becomes large, so we have \be \label{three}
{\epsilon^{5/4}\over g_s} \, \sim \, {N \epsilon^{1/4}} \, \to \,
0 \, . \ee Upon taking this limit, the dynamics of the $N$
D-particles reduces to the $c=1$ matrix quantum mechanics of the
tachyon mode $\hat{\TTT}$. Note that the limit in particular
involves sending the string coupling to zero (even fast enough so
that $g_s N \to 0$), so that the D-particles indeed decouple from
the IIB closed strings in the bulk.

\bigskip

\newsubsection{Conclusions}
\newcommand{\xx}{{\mbox{\large $x$}}}
\newcommand{\rr}{{\mbox{\large $r$}}}

Our new proposal is that 1+1-dimensional string theory, via its equivalence with
the $c=1$ matrix model, can be identified with the above-defined decoupling limit
of a dense collection of unstable D-particles in IIB string theory. The evidence
supporting this identification is twofold: (i) the worldline theory of the $N$ unstable
D-particles in this limit reduces to the $c=1$ matrix model, and (ii) the
interaction of the matrix model degrees of freedom with the 1+1-d closed strings is
consistent with their interpretation as the tachyon field of the corresponding
D-particles. We will now make some comments about this duality.

\medskip

\noindent (i) A perhaps somewhat surprising aspect of the proposed
duality is that it involves a dimensional reduction from a 10-d to
a 2-d string theory. This dimensional reduction amounts to the
statement that in the near-horizon limit only the $s$-wave sector
of the IIB string theory survives. A somewhat schematic
explanation for how this may come about is as follows. Consider
the full 10-d background produced by a dense collection of many
unstable IIB D-particles. This background has $SO(9)$ symmetry,
and is thus naturally describes in polar coordinates
$(\xx^0,\rr,\Omega)$. Since the boundary state of the D-particles
contains a tadpole for the dilaton and graviton, the worldsheet
action of a string moving in this background will take the general
form (omitting worldsheet fermions) \be S_{ws}= {1 \over 4 \pi}
\int d^2 \sigma \left(-{\rm A}(r)(\partial_a \xx^0)^2 +
(\partial_a \rr)^2 + \Phi(r) R^{(2)} + {\rm B}(\rr) (\partial_a
\Omega)^2 \right). \ee Now because of the non-trivial radial
dependence of the dilaton field $\Phi(\rr)$, the radial coordinate
$\rr$ acquires a non-trivial transformation under worldsheet scale
transformations $z \to (1+\delta \epsilon)z$: \be \delta \rr =
\delta \epsilon \, \nabla_r \Phi(\rr). \ee Thus if this gradient
gets large, we can interpret the radial evolution as a worldsheet
renormalization group flow: large values of $\rr$ correspond to
the ultra-violet, and small values of $\rr$ to the infra-red
region of the worldsheet CFT. Now if we freeze $\rr$, the angular
part of the worldsheet CFT reduces to an O(9) sigma-model with a
non-trivial RG flow: it is expected to become strongly coupled and
develop a mass-gap in the IR. In geometrical terms, this means
that the $S^8$ shrinks to zero size, leaving behind only the
$s$-wave modes. Unlike their supersymmetric IIA cousins, the IIB
D-particles do not produce a stabilizing flux through the $S^8$
that would prevent it from collapsing. Assuming that
the remaining radial and time coordinates remain massless, their
sigma-model action must be characterized by a solution to the 2-d
target space equation of motion of (\ref{target}).

\medskip

\noindent
(ii) It is natural to ask whether there are lessons from
all this about 26-dimensional critical string theory. If our
identification between the closed-string tachyon with the density
of D-particles persists in the higher dimensional theory, it
provides additional support for the long-standing suspicion
that the endpoint of the closed bosonic string
tachyon condensation may be described by 1+1-d non-critical string
theory \cite{workinprogress}.

\newcommand{\Aa}{{\mbox{\small $A$}}}

\medskip

\noindent (iii) The strategy followed in section 7 may possibly be
generalized to construct decoupling limits of many other unstable
D-brane systems, while reducing their world volume theory to
matrix quantum mechanics. The general idea is as follows.

Unstable D-branes generally arise as D-sphalerons, the minimal
energy configuration at the half-way point of a Euclidean
D-instanton trajectory. The D-instantons couple to some specific
RR-form, $C_{RR}$, which acts as the corresponding theta-angle. By
turning on a time-like gradient for $C_{RR}$ one can in effect
introduce a chemical potential that drastically reduces the height
of the sphaleron barrier, thus making the unstable D-branes very
light. By considering a suitable large $N$ limit, one may thus
hope to isolate a single matrix valued sphaleron mode.

As a concrete example of this procedure, consider Yang-Mills gauge
theory on $S^3\times R$. This theory has sphaleron configurations,
that for ${\cal N}=4$ SYM are related via AdS/CFT duality to
unstable D-particles in AdS \cite{sunny}. We can turn on a
chemical potential that makes the sphaleron mode light by
introducing a time-dependent theta-angle $\theta = \nu \, t$. This
results in an extra contribution to the SYM Hamiltonian
proportional to $\nu$ times the Chern-Simons number $CS = \int{\rm
Tr}(AdA + {2\over 3} A^3)$. (Such a modification of the
Hamiltonian naturally arises if one couples the gauge theory to
chiral fermions and considers the model at finite fermion density.
The parameter $\nu$ then corresponds to the chemical potential for
the anomalous fermion number $Q_f$.) In the limit where $\nu$
approaches a critical value $\nu_c$, the Hamiltonian for the
(appropriately defined \cite{vanBaal:1992xj}) constant gauge field
modes reduces to \be H \simeq {\rm Tr} \Bigl(E_i^2 \, +\,
\epsilon\, \Aa_i \Aa_i + g \epsilon^{ijk} \Aa_i [\Aa_j,\Aa_k]\,
\Bigr) \ee with $\epsilon = \nu_c -\nu \to 0$. In this limit, the
constant gauge field modes decouple from all the other modes. This
therefore defines a matrix model reduction of gauge theory. The
matrix potential has one unstable sphaleron mode, interacting with
two stable modes.

\medskip

\noindent (iv) The identification of 1+1-d closed strings, as
excitations of the Fermi sea of eigenvalues of the non-abelian
open string tachyon, opens up a very interesting new perspective
on the interactions between D-particles and strings. Whereas small
fluctuations of the Fermi sea have a direct perturbative closed
string interpretation, larger non-linear excitations may generate
configurations that can only be given a non-perturbative
interpretation. There are two distinct kinds of large
nonlinearities on the Fermi sea: large gradients and large
amplitudes.

When an incoming pulse has a spatial gradient above a certain
threshold, it will lead to a cresting, or folding, of the Fermi
surface \cite{Polchinski:uq}\cite{Polchinski:1994mb}\cite{Das:1995gd}. This
phenomenon, which has always been a somewhat puzzling feature of
the $c=1$ matrix model, now gets a rather natural interpretation
as the creation of an unstable D-particle from closed strings. The
life-time of the D-particle excitation grows logarithmically with
the inverse distance between the top of the wave and the top of
the potential. Eventually it decays into closed string
radiation. At intermediate times, however, it should reveal the
presence of on-shell {\it open} string excitations, which in the matrix
model correspond to excitations of size $\sqrt \hbar$ rather than
$\hbar$.

\figuur{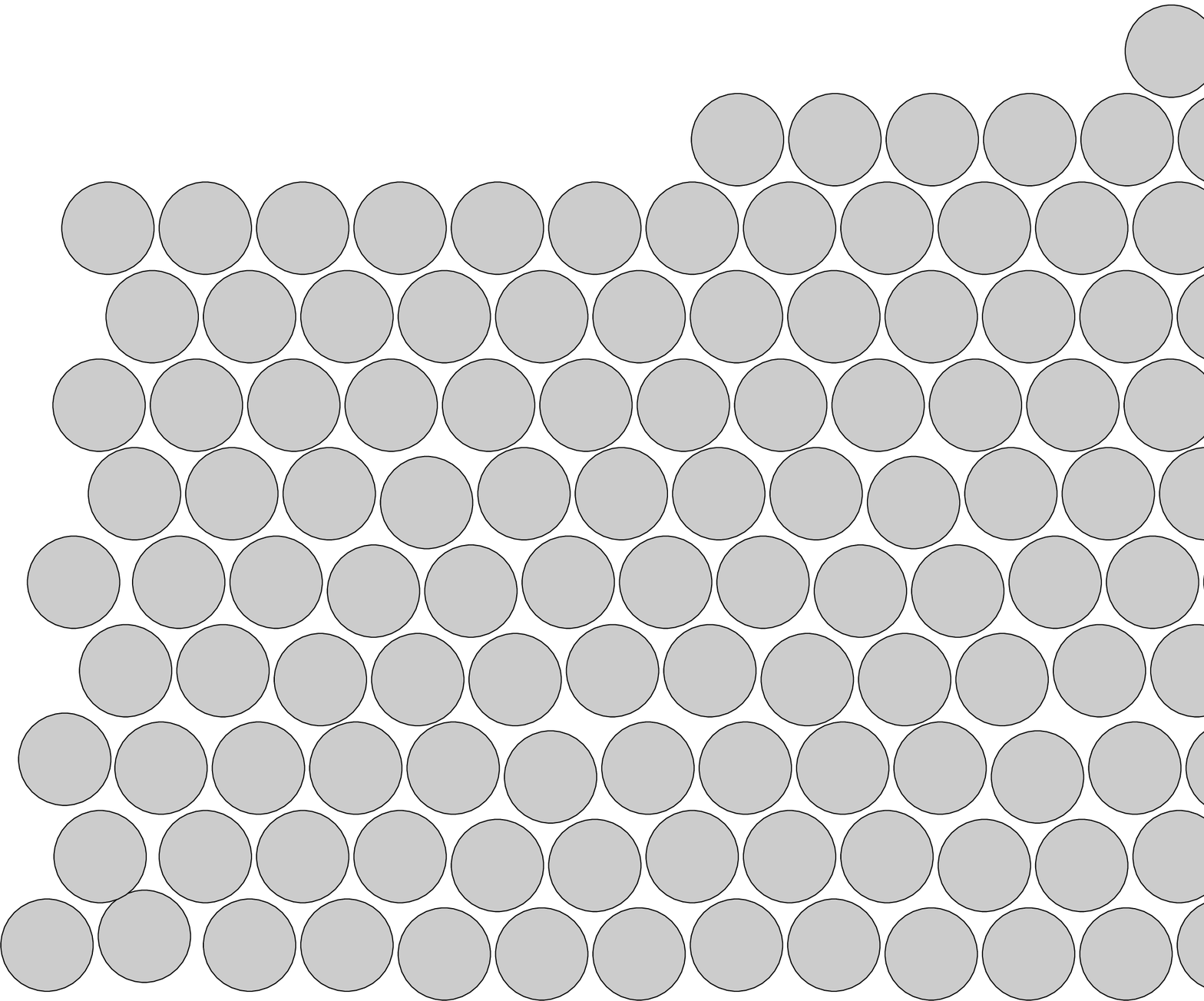}{6.5cm} {\it Fig.4 Cartoon of open string creation
during cresting of the fermi surface. Each circle represents an
eigenvalue occupying phase-space area $\hbar$.  The open
string depicted naturally has excitations of energy
$\sqrt \hbar$.}

The other kind of strong nonlinearity involves large amplitudes.
Again there is a threshold, which is when the pulse extends above
the top of the potential barrier.  In this case, there will be
tearing of the Fermi surface as the top of the wave is sucked into
the gorge of eternal peril. Luckily, via our new physical
interpretation, we can now clarify what happens to the part of the
wave that is lost: it corresponds to a collection of IIB D-particles
that decays via a tachyon that rolls down towards the right-hand side of the
effective potential in fig 2. The same fate awaits any eigenvalue
that penetrates the potential barrier via quantum mechanical
tunneling. The near-horizon $c=1$ string theory is therefore not a completely
decoupled theory, but nonetheless -- when thought
of as embedded inside of the
IIB string theory -- gives a completely adequate description of the physics of the
meta-stable bound state of D-particles.\footnote{One could perhaps try, however, to
obtain a self-consistent matrix model in which the Fermi sea is filled on both
sides of the potential barrier \cite{Dhar:1995gw}. In the IIB string theory, this
would need to be interpreted as a dense gas of both sD-branes and anti-sD-banes,
with positive and negative (\ref{cflux}) respectively.}

It is natural to look for nonlinear phenomena in the matrix model that can
be interpreted as black hole formation in the 1+1-d effective 
theory \cite{workinprogress}.
It will be interesting to revisit
the matrix black hole construction of
\cite{Kazakov:2000pm},
which can be interpreted as
turning on a Wilson line
of the D-particle world-line gauge field.

\medskip
\def\appendix#1{\addtocounter{section}{1}
\renewcommand{\thesection}{\Alph{section}}
\section*{Appendix \thesection\protect\indent \parbox[t]{11.15cm}
{#1} }
\addcontentsline{toc}{section}{Appendix \thesection\ \ \ #1}}

\bigskip
\bigskip

\noindent
{\large \bf Acknowledgements}

\noindent
We are grateful to Sunny Itzhaki for collaboration on the material in
section 7.
We would also like to thank
Allan Adams, Sujay Ashok, Davide Gaiotto, Simeon Hellerman,
Igor Klebanov,
David Kutasov, Hong Liu, Liat Maoz, Lubos Motl, Sameer Murthy, Sasha Polyakov,
Leonardo Rastelli, and Spenta Wadia for discussions.
The work of JM is supported by a Princeton University Dicke Fellowship.
This work is supported by the National Science
Foundation under Grant No. 98-02484.
Any opinions, findings, and conclusions or recommendations expressed in
this material are those of the authors and do not necessarily reflect
the views of the National Science Foundation.

\bigskip
\bigskip

\appendix{}

In this Appendix we will compute the closed string tachyon state produced by
a rolling open string tachyon on a single D-particle. We will do the
calculation for the critical bounce trajectory (\ref{bounce}) with
$\lambda={1\over 2}$, though our result is easily generalized to arbitrary
$\lambda \leq {1\over 2}$.
We will find that the rolling tachyon produces, apart from a discrete series of
interesting transients,  also a {\it static} shift in the closed string tachyon,
which  needs to be interpreted as a small shift in $\mu$. The result will
be in accord with our proposed duality between the D-particles and
the $c=1$ matrix eigenvalues.  We will also see that the $1+1$-d string theory
gives an extremely clean example of the paradigm for describing closed strings as
imaginary D-branes, recently advocated in \cite{stherm}\cite{Lambert:2003zr}
\cite{GIR}.

We will extract the response of the closed-string tachyon to
the D-particle by factorizing the annulus diagram onto on-shell
closed-string poles.
We begin by considering the annulus amplitude
\be
\CA_{s_1, s_2}(\omega)=
\int\limits_0^1 \!\! d\tilde q\,
\bra{B_{s_2}} \otimes \langle\bra{\omega} ~\tilde q^{L_0}
\ket{ B_\lambda } \otimes \ket{ B_{s_1}}
\cdot \CA_{\rm ghost} .
\ee
Here $\langle\bra{\omega} $ is the Ishibashi
state built on the primary $ \bra{ 0}\, e^{- i \omega X^0} $
of the $X^0$ CFT.
Note that we have already performed the integral over the
phase of the closed string modular parameter $\tilde q$, which
implements level-matching, $L_0 = \bar L_0$.
The ghost part of the annulus is
\be
\CA_{\rm ghost } = {\eta^2(\tilde q) \over \tilde q}.
\ee
The $X^0$ piece of this amplitude is
\be
\langle\bra{\omega} ~\tilde q^{L_0}
\ket{ B_\lambda }
=  {\vev{e^{-i \omega X^0}}\over \eta(\tilde q)}
= {\pi \, \tilde{q}^{-\omega^2}
\over  \eta(\tilde q) \sinh \pi \omega}
\ee
where the expectation value denotes the one-point function on the disk.
Finally, the Liouville part of this amplitude is \cite{Teschner:2000md}
\be
\int_{\CC} {dP \over 2 \pi i} ~{\tilde q^{P^2} \over \eta(\tilde q)}
~ \vev{ B_{s_2} | v_P } \, \vev{ v_P | B_{s_1} }
\ee
where the matrix elements are as given in (\ref{fuzzy}).
The contour of integration is $ \CC = - i {Q\over 2} + \IR $.

Putting things together, we find that all factors $\eta(\tilde q)$ cancel,
leaving a trivial integration over the modular parameter $\tilde q$.
It reduces to the massless propagator
\be \int\limits_0^1 \! {d \tilde q \over\tilde q} ~\tilde q^{P^2-\omega^2} = { 1
\over P^2-\omega^2} .
\ee
To extract the resulting on-shell background, we follow the general prescription
derived in \cite{GIR} and take the discontinuity in $\omega$ of this object:
\be
{1 \over P^2 - \omega^2 }
\longrightarrow
i \pi {\delta (P-\omega) \over \omega} ,
\ee
so that we can use the delta-function to do the integral over $P$.
Collecting all these facts, we have
\be
\CA_{s_1, s_2}(\omega) =
\tilde c {i \pi \cos (\pi s_2 \omega) \cos(\pi s_1 \omega) \over (\sinh
\pi \omega)^3}, \ee

Now let us specialize the boundary states to fit the physical
problem we wish to study. First we set $s_1={i \over 2}$ so that
the state $\ket{B_{s_1}}$ represents the D-particle. The other
state we decompose as \be \bra{B_{s_2}} = \int \! {d\ell \over
\ell} \, e^{-\ell \sqrt{\mu} \cosh\pi s_2} \, \bra{W(\ell)}\, .
\ee The states $\bra{W(\ell)}$ we interpret as the position
eigenstate for the tachyon mode via the identification $\ell =
e^{\varphi}$. Using once again the integral formula (\ref{geez}),
we can thus write the amplitude as \be \CA_{s_1, s_2} = i \tilde c
\int_0^\infty {d\ell \over \ell}\, e^{- \ell \sqrt \mu \cosh \pi
s_2} { \pi K_{i\omega} (\sqrt \mu
\, \ell) 
\over \sinh (\pi \omega)\, \sinh(\pi \omega/2)}
\ee
The position-space profile
of the shift in the closed string tachyon is therefore
\be \delta {\cal T}(\ell, \omega) = \tilde c { \pi\, \ell^2 \,
K_{i\omega} (\sqrt \mu \, \ell) \over \sinh(\pi \omega)\,
\sinh(\pi\omega / 2) } \ee To extract the time-dependence of this
background, we Fourier transform using the contour in the figure;
the integral may be done by residues.

\figuur{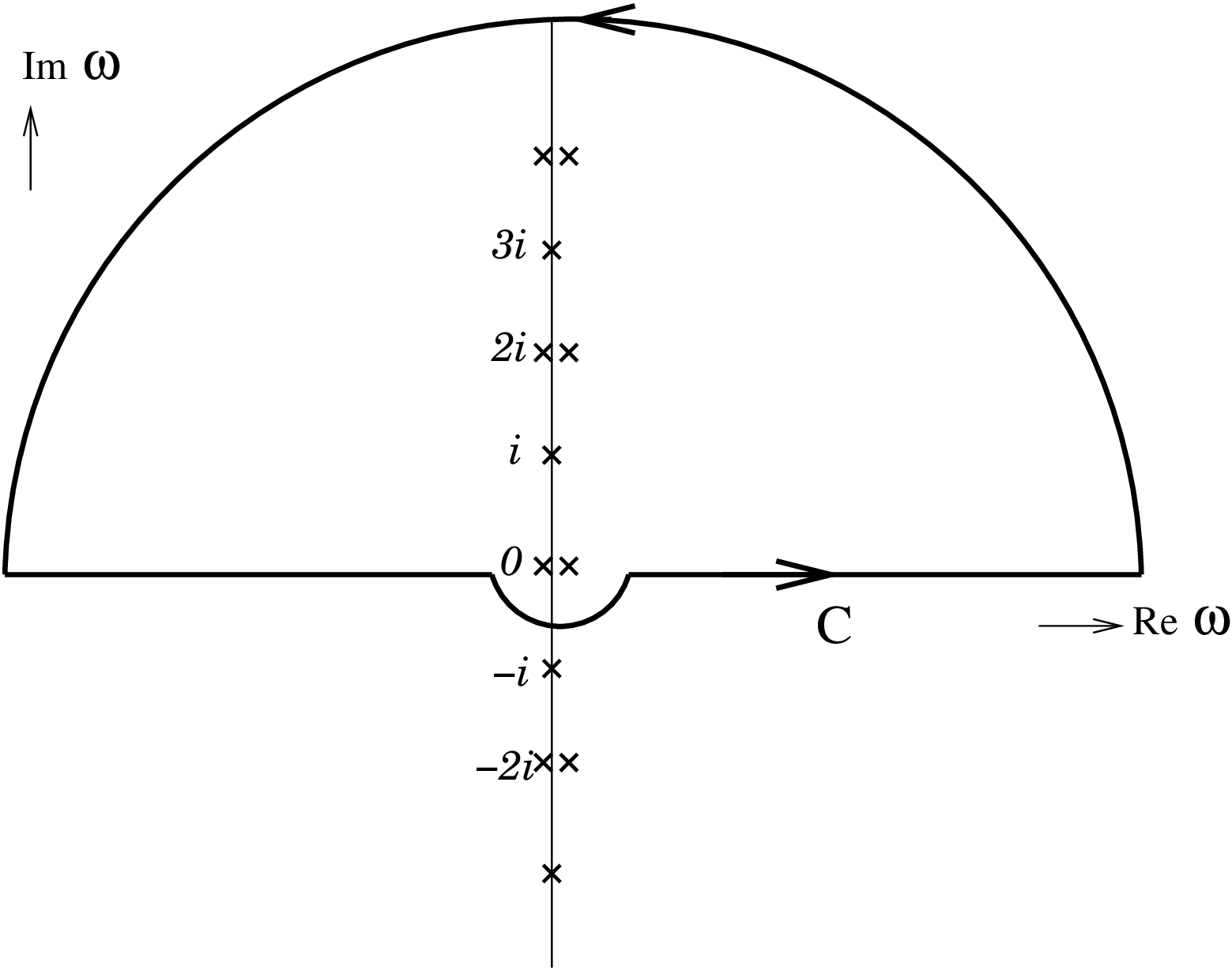}{8cm}{Fig 3.
The denominator
$$\sinh \pi \omega ~\sinh { \pi \omega \over 2}
= \pi^2 \omega^2 \prod_{n = 1}^\infty
\left( \left( \omega \over 2n \right)^2 + 1 \right)^2
 \left(
\left( \omega \over 2n+1 \right) ^2 + 1
\right)
$$
has double zeros at even imaginary integers and
single zeros at odd imaginary integers.  The physical
tachyon response is obtained by Fourier transforming
using this contour.}
\be
\delta {\cal T}(\ell, t)
= \int {d\omega \over 2 \pi}\, e^{ i \omega t} \delta T(\ell, \omega)
= \delta {\cal T}(\ell) +
\sum_{ n = 1}^\infty
e^{- n t} c_n(\ell)
\ee
The terms with finite $n$ are transients which
represent the
splash of the probe D-particle into the Fermi sea.
The momenta of the transients are quantized in units of the
frequency of the harmonic oscillator
appearing in the euclidean continuation of the
matrix quantum mechanics.  The distinction between
odd and even multiples of this basic frequency
is the distinction between lengths and laps
under the barrier.

The static piece of the shift in the tachyon background is (with
$\ell = e^\varphi$) \be \delta {\cal T}(\varphi) \propto  \ell^2
\left.
\partial_\nu K_{\nu} ( \sqrt \mu \, \ell) \right|_{\nu = 0} =
\ell^2 I_0 (\sqrt \mu \ell). \ee which in the asymptotic region
amounts to a shift $\delta {\cal T}(\varphi) \propto e^{2
\varphi}$. Hence the shift at the location $\varphi =0$ of the
``dilaton wall'' is of order one. Since ${\cal T}(0) \simeq
{\mu\over 2} \log \mu$, we find that the presence of the extra
D-particle amounts to a shift $\delta \mu$ of order \be \delta \mu
\propto ( \log \mu)^{-1} , \ee in accordance with the
characteristic level density $\rho(\mu) = {\partial N\over
\partial \mu} \simeq -{2\over \pi} \log \mu$  of the $c=1$ matrix
model.




\begin{thebibliography}{99}



\bibitem{Sen:99}
A.~Sen, ``Non-BPS states and branes in string theory,''
arXiv:hep-th/9904207.



\bibitem{piljin1}
P.~Yi,
``Membranes from five-branes and fundamental strings from Dp branes,''
Nucl.\ Phys.\ B {\bf 550}, 214 (1999)
[arXiv:hep-th/9901159];
O.~Bergman, K.~Hori and P.~Yi,
``Confinement on the brane,''
Nucl.\ Phys.\ B {\bf 580}, 289 (2000)
[arXiv:hep-th/0002223].









\bibitem{Sen:2002nu}
A.~Sen, ``Rolling tachyon,'' JHEP {\bf 0204}, 048 (2002)
[arXiv:hep-th/0203211];
``Tachyon matter,'' JHEP {\bf 0207}, 065 (2002)
[arXiv:hep-th/0203265];
``Field theory of tachyon matter,'' Mod.\ Phys.\ Lett.\ A {\bf
17}, 1797 (2002) [arXiv:hep-th/0204143].


\bibitem{Sen:2002vv}
A.~Sen, ``Time evolution in open string theory,'' JHEP {\bf 0210},
003 (2002) [arXiv:hep-th/0207105];
``Time and tachyon,'' arXiv:hep-th/0209122.


\bibitem{Gutperle:2002ai}
M.~Gutperle and A.~Strominger, ``Spacelike branes,'' JHEP {\bf
0204}, 018 (2002) [arXiv:hep-th/0202210]; A.~Strominger, ``Open
string creation by S-branes,'' arXiv:hep-th/0209090.

\bibitem{Gutperle:2003xf}
M.~Gutperle and A.~Strominger, ``Timelike boundary Liouville
theory,'' arXiv:hep-th/0301038.




\bibitem{stherm}
A.~Maloney, A.~Strominger and X.~Yin, ``S-brane thermodynamics,''
arXiv:hep-th/0302146.





\bibitem{Lambert:2003zr}
N.~Lambert, H.~Liu and J.~Maldacena, ``Closed strings from
decaying D-branes,'' arXiv:hep-th/0303139.



\bibitem{GIR}
D.~Gaiotto, N.~Itzhaki and L.~Rastelli, ``Closed Strings as
Imaginary D-branes,'' arXiv:hep-th/0304192.


\bibitem{longlist}
E.~Brezin, C.~Itzykson, G.~Parisi and J.~B.~Zuber,
``Planar Diagrams,''
Commun.\ Math.\ Phys.\  {\bf 59}, 35 (1978).


\bibitem{Ginsparg:is}
P.~Ginsparg
and G.~W.~Moore, ``Lectures On 2-D Gravity And 2-D String
Theory,'' arXiv:hep-th/9304011.



\bibitem{Klebanov:1991qa}
I.~R.~Klebanov, ``String theory in two-dimensions,''
arXiv:hep-th/9108019.




\bibitem{Polchinski:1994mb}
J.~Polchinski, ``What is string theory?,'' arXiv:hep-th/9411028.


\bibitem{Shenker:1990uf}
S.~H.~Shenker,
``The Strength Of Nonperturbative Effects In String Theory,''
RU-90-47
{\it Presented at the Cargese Workshop on Random Surfaces, Quantum Gravity and Strings, Cargese, France, May 28 - Jun 1, 1990}






\bibitem{matrix}
A. Wachowski and L. Wachowski,
``The Matrix: Reloaded,''
to appear.


\bibitem{Polchinski:1994jp}
J.~Polchinski, ``On the nonperturbative consistency of d = 2
string theory,'' Phys.\ Rev.\ Lett.\  {\bf 74}, 638 (1995)
[arXiv:hep-th/9409168].




\bibitem{Polyakov:2001af}
A.~M.~Polyakov,
``Gauge fields and space-time,''
Int.\ J.\ Mod.\ Phys.\ A {\bf 17S1}, 119 (2002)
[arXiv:hep-th/0110196].



\bibitem{piljin2}
G.~Gibbons, K.~Hashimoto and P.~Yi,
``Tachyon condensates, Carrollian contraction of Lorentz group, and  f\
undamental strings,''
JHEP {\bf 0209}, 061 (2002)
[arXiv:hep-th/0209034];
G.~W.~Gibbons, K.~Hori and P.~Yi,
``String fluid from unstable D-branes,''
Nucl.\ Phys.\ B {\bf 596}, 136 (2001)
[arXiv:hep-th/0009061].



\bibitem{Okuda:2002yd}
T.~Okuda and S.~Sugimoto, ``Coupling of rolling tachyon to closed
strings,'' Nucl.\ Phys.\ B {\bf 647}, 101 (2002)
[arXiv:hep-th/0208196].

\bibitem{Leblond:2003db}
F.~Leblond and A.~W.~Peet,
arXiv:hep-th/0303035.


\bibitem{kutasov:2003er}
D.~Kutasov and V.~Niarchos,
``Tachyon effective actions in open string theory,''
arXiv:hep-th/0304045.






\bibitem{Polchinski:mf}
J.~Polchinski,
``Critical Behavior Of Random Surfaces In One-Dimension,''
Nucl.\ Phys.\ B {\bf 346}, 253 (1990).


\bibitem{Seiberg:1990eb}
N.~Seiberg,
``Notes On Quantum Liouville Theory And Quantum Gravity,''
Prog.\ Theor.\ Phys.\ Suppl.\  {\bf 102}, 319 (1990).


\bibitem{Polchinski:1990mh}
J.~Polchinski,
``Remarks On The Liouville Field Theory,''
UTTG-19-90
{\it Presented at Strings '90 Conf., College Station, TX, Mar 12-17, 1990}





\bibitem{Fateev:2000ik}
V.~Fateev, A.~B.~Zamolodchikov and A.~B.~Zamolodchikov,
``Boundary Liouville field theory. I: Boundary state and boundary  two-point function,''
arXiv:hep-th/0001012.

\bibitem{Teschner:2000md}
J.~Teschner,
``Remarks on Liouville theory with boundary,''
arXiv:hep-th/0009138.






\bibitem{Zamolodchikov:2001ah}
A.~B.~Zamolodchikov and A.~B.~Zamolodchikov,
``Liouville field theory on a pseudosphere,''
arXiv:hep-th/0101152.

\bibitem{emil}
E.~J.~Martinec,
``The annular report on non-critical string theory,''
arXiv:hep-th/0305148.


\bibitem{nowwehaveexactlywhatwewant}
I.~Klebanov, J.~Maldacena, N.~Seiberg,
hep-th/0305159.

\bibitem{joerg}
J.~McGreevy, J.~Teschner, H.~Verlinde,
hep-th/0305194.





\bibitem{Rajaraman:1999hn}
A.~Rajaraman and M.~Rozali,
``D-branes in linear dilaton backgrounds,''
JHEP {\bf 9912}, 005 (1999)
[arXiv:hep-th/9909017].

\bibitem{Ponsot:2001ng}
B.~Ponsot and J.~Teschner,
``Boundary Liouville field theory: Boundary three point function,''
Nucl.\ Phys.\ B {\bf 622}, 309 (2002)
[arXiv:hep-th/0110244].



\bibitem{Kostov:2002uq}
I.~K.~Kostov,
``Boundary correlators in 2D quantum gravity: Liouville versus discrete  approach,''
arXiv:hep-th/0212194.



\bibitem{Moore:1991ir}
G.~W.~Moore, N.~Seiberg and M.~Staudacher,
``From loops to states in 2-D quantum gravity,''
Nucl.\ Phys.\ B {\bf 362}, 665 (1991).



\bibitem{Kazakov:1991pt}
V.~A.~Kazakov and I.~K.~Kostov,
``Loop gas model for open strings,''
Nucl.\ Phys.\ B {\bf 386}, 520 (1992)
[arXiv:hep-th/9205059].




\bibitem{dj}
S.~R.~Das and A.~Jevicki, ``String Field Theory And Physical
Interpretation Of D = 1 Strings,'' Mod.\ Phys.\ Lett.\ A {\bf 5},
1639 (1990).



\bibitem{Polchinski:fq}
J.~Polchinski,
``Combinatorics Of Boundaries In String Theory,''
Phys.\ Rev.\ D {\bf 50}, 6041 (1994)
[arXiv:hep-th/9407031].




\bibitem{Dhar:1995gw}
A.~Dhar, G.~Mandal and S.~R.~Wadia,
``Discrete state moduli of string theory from the C=1 matrix model,''
Nucl.\ Phys.\ B {\bf 454}, 541 (1995)
[arXiv:hep-th/9507041].








\bibitem{Minahan:1992bz}
J.~A.~Minahan,
Int.\ J.\ Mod.\ Phys.\ A {\bf 8}, 3599 (1993)
[arXiv:hep-th/9204013].



\bibitem{Yang:hs}
Z.~Yang,
``Fermion Coupling And Dynamical Loops In D = 1 Random Matrix Models,''
UTTG-20-90.










\bibitem{Garousi:2000tr}
M.~R.~Garousi,
``Tachyon couplings on non-BPS D-branes and
Dirac-Born-Infeld action,'' Nucl.\ Phys.\ B {\bf 584}, 284 (2000)
[arXiv:hep-th/0003122].
\bibitem{Kluson:2000iy}
J.~Kluson,
``Proposal for non-BPS D-brane action,'' Phys.\ Rev.\ D
{\bf 62}, 126003 (2000) [arXiv:hep-th/0004106].
\bibitem{Bergshoeff:2000dq}
E.~A.~Bergshoeff, M.~de Roo, T.~C.~de Wit, E.~Eyras and S.~Panda,
``T-duality and actions for non-BPS D-branes,'' JHEP {\bf 0005},
009 (2000) [arXiv:hep-th/0003221].




\bibitem{Okuyama:2003wm}
K.~Okuyama,
``Wess-Zumino Term in Tachyon Effective Action,''
arXiv:hep-th/0304108.

\bibitem{Sen:2003tm}
A.~Sen, ``Dirac-Born-Infeld action on the tachyon kink and
vortex,'' arXiv:hep-th/0303057.









\bibitem{workinprogress}
Work in progress.

\bibitem{sunny}
N.~Drukker, D.~J.~Gross and N.~Itzhaki,
``Sphalerons, merons and unstable branes in AdS,''
Phys.\ Rev.\ D {\bf 62}, 086007 (2000)
[arXiv:hep-th/0004131].




\bibitem{vanBaal:1992xj}
P.~van Baal and N.~D.~Hari Dass, ``The Theta dependence beyond
steepest descent,'' Nucl.\ Phys.\ B {\bf 385}, 185 (1992);
B.~van den Heuvel, ``Glueballs on the three-sphere,'' Nucl.\
Phys.\ B {\bf 488}, 282 (1997) [arXiv:hep-lat/9608101].






\bibitem{Polchinski:uq}
J.~Polchinski,
``Classical Limit Of (1+1)-Dimensional String Theory,''
Nucl.\ Phys.\ B {\bf 362}, 125 (1991);
M.~Natsuume and J.~Polchinski,
``Gravitational Scattering In The C = 1 Matrix Model,''
Nucl.\ Phys.\ B {\bf 424}, 137 (1994)
[arXiv:hep-th/9402156].












\bibitem{Das:1995gd}
S.~R.~Das and S.~D.~Mathur,
``Folds, bosonization and nontriviality of the classical limit of 2-D string theory,''
Phys.\ Lett.\ B {\bf 365}, 79 (1996)
[arXiv:hep-th/9507141].



\bibitem{Kazakov:2000pm}
V.~Kazakov, I.~K.~Kostov and D.~Kutasov,
``A matrix model for the two-dimensional black hole,''
Nucl.\ Phys.\ B {\bf 622}, 141 (2002)
[arXiv:hep-th/0101011].


\bibitem{Adams:2001jb}
A.~Adams and E.~Silverstein,
``Closed string tachyons, AdS/CFT, and large N QCD,''
Phys.\ Rev.\ D {\bf 64}, 086001 (2001)
[arXiv:hep-th/0103220].






\end{thebibliography}
\end{document}